\documentclass[3p,times,11pt]{elsarticle}

\usepackage{lineno,hyperref}
\usepackage{amsmath}
\usepackage{graphicx}
\usepackage{subcaption}
\usepackage{adjustbox}
\usepackage{color}

\modulolinenumbers[5]

\journal{Journal of Computational Physics}

\begin{document}

\begin{frontmatter}

\title{Variational Multi-scale Super-resolution : A data-driven approach for reconstruction and predictive modeling of unresolved physics.}

\author{Aniruddhe Pradhan}
\ead{anipra@umich.edu}
\author{Karthik Duraisamy}
\ead{kdur@umich.edu}
\address{Department of Aerospace Engineering, University of Michigan, Ann Arbor, MI 48109, USA}

\begin{abstract}
The variational multiscale (VMS) formulation formally segregates the evolution of the coarse-scales from the fine-scales. VMS modeling requires the approximation of the impact of the fine scales in terms of the coarse scales. {\color{black} In linear problems,  our formulation reduces the problem of learning the sub-scales  to learning the projected element Green's function basis coefficients. For the purpose of this approximation,  a special neural-network structure - the variational super-resolution N-N (VSRNN) - is proposed.} The VSRNN constructs a super-resolved model of the unresolved scales as a sum of the products of individual functions of coarse scales and physics-informed parameters. Combined with a set of locally non-dimensional features obtained by normalizing the input coarse-scale and output sub-scale basis coefficients, the VSRNN provides a general framework for the discovery of closures for both the continuous and the discontinuous Galerkin  discretizations. {\color{black}By training this model on a sequence of $L_2-$projected data and using the subscale to compute the continuous Galerkin subgrid terms, and the super-resolved state to compute the discontinuous Galerkin fluxes, we improve the optimality and the accuracy of these methods for the convection-diffusion problem, linear advection and turbulent channel flow. Finally, we demonstrate that - in the investigated examples - the present model allows generalization to out-of-sample initial conditions and Reynolds numbers}. Perspectives are provided on data-driven closure modeling, limitations of the present approach, and opportunities for improvement.
\end{abstract}

\begin{keyword}
\texttt{Variational Multiscale Method, Super-resolution, Physics-informed Deep Learning, Coarse-grained Modeling, Continuous Galerkin, Discontinuous Galerkin}
\end{keyword}

\end{frontmatter}

\section{Introduction}
Multi-scale problems are ubiquitous in science and engineering, and many practical applications are characterized not just by a disparity of scales, but also by a complex interplay between scales. 
The coarse scales which are resolved in a multi-scale computation constitute the macro-scale level of description. The finer, unresolved scales constitute the micro-scale level of description, and their effect on the coarse-scales needs to be modeled not just for accuracy, but also for stability~\cite{VMS}.
A classical application of multi-scale modeling is turbulent fluid-flow at high Reynolds numbers. 
 In the so-called Large Eddy Simulations (LES) of turbulent flow, the energy-containing large scales are resolved and the effect of the small scales, also known as sub-scales, are modeled~\cite{DSM,GDSM,DSM2,SIGMA,VREMEN,WALE,OSS,OSS2,VMS3,VMSE,NLVMS,NLVMS2,MZVMS}. 

Over the past few decades, several modeling approaches have been pursued to improve the performance of coarse-grained simulation of multi-scale PDEs. These techniques can be broadly classified into two categories: (i.) physics-inspired approaches that employ phenomenological assumptions; and (ii.) structural formulations that attempt to derive {\color{black}subgrid} models from the structure of the governing PDE. Since it is not possible to review all such approaches, we highlight representative examples in the area of turbulence simulations, and use the discussion to set the context of the present work.

Physics-based models use arguments such as the balance of energy transfer from large to small scales or scale-similarity (e.g. the Smagorinsky\cite{SMAG} {\color{black}model} in LES and its adaptive variants~\cite{DSM,lilly1992proposed,DSM2}.)  
Additional functional forms were proposed by Vreman~\cite{VREMEN}, Nicoud et al. \cite{WALE,SIGMA}, and L{\'e}v{\^e}que et al. \cite{leveque2007shear} to correct asymptotic near-wall behavior. Another class of subgrid models based on the self-similarity idea was first put forward by Bardina et al. \cite{bardina1980improved} in its original form, and later explored by other researchers in the mixed form \cite{bardina1980improved, liu1994properties, liu1995experimental, vreman1994formulation, xie2019artificial}. The self-similarity model in its original form was found to have a high spatial correlation with the actual sub-grid stress but lacked enough dissipation to ensure the stability of the approach. 

Langford et al. \cite{optimal} in their work on optimal-LES showed that it is possible to construct an abstract sub-grid model that can obtain correct single-time, multi-point statistics and generates  minimal error in the instantaneous dynamics. This is possible by minimizing the root mean square error in the time derivative of the resolved quantities. Here, the LES model is written in terms of a conditional average, an average over all the instantaneous fields that correspond to the same LES solution when filtered. As a result, the modeled sub-grid stresses {\color{black}do} not necessarily have to show a high spatial correlation with the true sub-grid stresses. In addition to  these modeling approaches, the interaction of the numerical method with the under-lying sub-grid model also poses several challenges. The operating filter size in all these models is, in general, very close to the grid size. Consequently, the truncation error due to the numerical method  can be of a similar magnitude to the sub-grid model term. To decouple them, various explicit filtering approaches \cite{bose2010grid,lund2003use,lund1995experiments} have also been explored in the literature. The usage of dissipative numerical methods alone to perform LES without any explicit sub-grid model has also gained popularity\cite{moura2017eddy,flad2017use,fernandez2017subgrid,hickel2006adaptive,sun2018implicit} recently in the form of implicit-LES.

Another category of sub-grid models is based on the mathematical structure of the PDEs. 
One such mathematical formalism is the variational multiscale (VMS) method \cite{VMS,hughes2005variational} which formally separates the evolution equation of the coarse scales from the evolution equation of the fine-scales. The final VMS modeling step involves writing an explicit expression for the fine-scales approximately in-terms of the coarse-scale residual. These methods \cite{VMS,GLS,SUPG,ADJ,SUPG2} were first developed to {\color{black}stabilize} finite element methods for linear problems involving advection but later extended to non-linear problems \cite{OSS2,OSS,VMS3,VMSE,NLVMS,NLVMS2} as sub-grid models. Analogous model forms for the sub-scales in terms of the coarse-scale residual can also be derived for the non-linear problems if the Mori-Zwanzig (M-Z) formalism is used with the VMS method\cite{MZ1,MZ2,MZVMS,pradhan2019variational,parish2016paradigm}. In this work, we attempt to discover such VMS model forms directly from data using deep learning tools, and use them for predictive modeling.

An alternate class of methods in structural sub-grid modelling is based on the approximate deconvolution method (ADM) \cite{stolz1999approximate,stolz2001approximate}. In the ADM approach, the filtered state variables are approximately deconvolved using ADM operators approximated by a truncated series expansion of the inverse filter. The non-linear terms in the governing equations are then computed using these deconvolved state variables.

The availability of high resolution data from numerical simulations and experiments in the past decade has led to an interest in  data-based modeling. Applications of machine learning augmentations have been used in RANS  (e.g.  \cite{singh2016using,parish2016paradigm,ling2016reynolds,wang2017physics} and LES (e.g. \cite{sarghini2003neural,gamahara2017searching,xie2019modeling,beck2019deep}). In the LES front, Maulik et al. \cite{maulik2019sub} used machine learning to classify and blend different LES models to select the most accurate model at run-time. Yang. et al \cite{yang2019predictive} used physics-informed features to improve the performance of equilibrium LES wall models in non-equilibrium cases. Similarly, many other notable attempts to improve LES models using data have also been made by Maulik et al. \cite{maulik2017neural, maulik2018data,maulik2019sub}, Beck et al. \cite{beck2019deep}, Sarghini et al. \cite{sarghini2003neural}, Ghamara and Hattori \cite{gamahara2017searching}, Wang et. al \cite{wang2018investigations} and many more \cite{xie2019modeling,xie2019artificial,xie2020modeling}. These data-driven techniques have also found application in developing closures for reduced-order models (ROMs) \cite{mou2021data,xie2018data,mohebujjaman2019physically,wang2020recurrent}. 

Very recently,  super-resolution of turbulent-flow fields has been pursued using neural networks \cite{xie2018tempogan,deng2019super,liu2020deep,fukami2019super,kim2020unsupervised,fukami2020machine}.  Xie et al. \cite{xie2018tempogan} and Fukami et al. \cite{fukami2019super} appear to be the first to introduce this idea in fluid dynamic applications. These were followed by Deng et al. \cite{deng2019super} who improved traditional GAN performance by augmenting the model architecture. Improvements in flow field reconstruction were shown by Liu et al. \cite{liu2020deep}, using both spatial and temporal information. Fukami et al. \cite{fukami2020machine} performed super-resolution and in-betweening to reconstruct a highly-resolved space-time solution using two low-resolution snapshots taken at the start $t$ and the end $t+\Delta t$ of an interval. These models have demonstrated an ability to reconstruct  fine scales from highly coarse-grained data for either the same or similar data-set on which they have been trained both in a supervised and an unsupervised setting \cite{kim2020unsupervised}. However, applying the trained models to super-resolve coarse flow-fields at different Reynolds numbers or another part of the flow,  is relatively unexplored. 

The idea of using the super-resolved field to compute the closure terms is similar to the ADM approach. However, compared to the approximately deconvolved solution that lies on the same mesh as the filtered solution, the super-resolved solution lies on a higher resolution mesh. A super-resolution model capable of reconstructing fine-space data for a case where the fine-space data already exists and the same information is used for training has no use in a predictive setting. This paper attempts to improve  predictive capabilities by bringing in generalizable model forms and features. 
 In addition to the generalizability of these models, the definition of a coarse-space solution is ambiguous. This ambiguity is because a variety of low-fidelity data, including LES solution, obtained using finite difference method (FDM), finite volume method (FVM), spectrally filtered DNS solution, and stabilized finite element (FE) solution on a coarse grid qualify as coarse-solutions. The nodal or modal values in each of these methods represent different quantities. For example, Fukami et al. \cite{fukami2019super} used the max-pooling operation to obtain coarse data. Consequently, the trained model is  dependent on the type of method or filter used to generate data, and the mapping learned by the network has no formal basis. To resolve this ambiguity, we define both our coarse space and our fine space in terms of the $L_2$ projection of the DNS solution on low and high order polynomial basis functions in a similar spirit to the Variational Multiscale Method. As a result, the trained model will approximate the function that maps the $L_2$ projection of the DNS data on the two sub-spaces. Additionally, the model should be preferably compact and applied patch-wise rather than on the entire flow. This is because there is no guarantee that the coarse data that needs to be super-resolved has the same size as that of the input layer, and interpolating it back to the network size defeats the purpose of super-resolution. In this work, we develop N-N closures that are: (i.) capable of extrapolating to unseen flow conditions and resolutions; (ii.) use non-dimensional features rather than dimensional features for better generalizability; and (iii.) can be applied patch-wise rather than on the entire field.  

The outline of the paper is as follows: We introduce the VMS methodology in section 2. In section 3, we derive VMS consistent features. In section 4, we propose a  model form and a new network architecture for learning it. We describe the procedure of generating training data in section 5. In section 6, we apply our approach to the linear advection problem both in an online (numerical method) and offline (super-resolution) setting. In section 7, we evaluate the performance of a model approach to the turbulent channel flow. Perspectives on the broader challenge of data driven modeling, and on the present work is shared in section 8. Finally, we summarize our work in section 9. 

\section{The Variational Multiscale (VMS) Method}
This section summarizes the Variational Multiscale Method (VMS), which was originally presented by Hughes et al.\cite{VMS}. As discussed previously, this method has been extensively used for developing closures for both linear \cite{VMS,GLS,SUPG,ADJ,SUPG2} and non-linear PDEs \cite{OSS2,OSS,VMS3,VMSE,NLVMS,NLVMS2}. This section will only discuss it in the context of a linear problem and use it as a guiding principle for feature selection and in shaping the network architecture. As discussed by Hughes et al.\cite{VMS}, the development of VMS closure can be broadly categorized into two different subsections based on the type of basis functions used which are detailed below, along with a context for super resolution. 

\subsection{Smooth Case}
In the ‘smooth case’, the basis functions are sufficiently smooth so that the distributional effects may be ignored \cite{VMS}. Both the Fourier basis and the Chebyshev spectral basis qualify as a smooth basis. For this case, consider the following PDE on an open and bounded domain $\Omega \subset \mathbb{R}^d$, where  $d\geq1$ is the dimension of the problem, with a smooth boundary $\Gamma = \partial \Omega$:
\begin{equation}
    \mathcal{L}(u) = f \quad in \quad \Omega , \ \ 
    u = g \quad on \quad \Gamma
    \label{mainPDE}
\end{equation}
where the operator $\mathcal{L}$ can be linear or non-linear, the functions $f:\Omega\rightarrow\mathbb{R}$ and $g:\Gamma\rightarrow\mathbb{R}$ are given. The variational form of the above PDE is given by
\begin{equation}
    (\mathcal{L}(u),w)=(f,w), 
\end{equation}
such that $u \in \mathcal{V}$ for all $w \in \mathcal{V}$, where $(\cdot,\cdot)$ denotes the $L_2(\Omega)$ inner product, and $\mathcal{V}\equiv\mathcal{H}^1(\Omega)$ is the Sobolev space. The solution and weighting space are decomposed as follows:
\begin{equation}
\mathcal{V} = {\mathcal{V}_h} \oplus \mathcal{V}',
\end{equation}
where $\oplus$ represents a direct sum of ${\mathcal{V}_h}$ and $\mathcal{V}'$. Applying the VMS operation, we have
\begin{equation}
    (\mathcal{L}(u_h+u'),w_h)+ (\mathcal{L}(u_h+u'),w') =(f,w_h)+(f,w').
\end{equation}
While the above equation is valid for both non-linear and linear equations, further simplifications can be made if the differential operator is assumed to be linear. To this end, using the linear independence of $w_h$ and $w'$, and taking the differential operator to be linear, we obtain the coarse and fine equations :
\begin{align}
    (\mathcal{L}(u_h),w_h) +  (\mathcal{L}(u'),w_h) &=  (f,w_h) \\
    (\mathcal{L}(u'),w') &= - (\mathcal{L}u_h-f,w').
\end{align}
The coarse and fine scale equations can be re-written as:
\begin{align}
    (\mathcal{L}(u_h)-f,w_h)  &=  -(\mathcal{L}(u'),w_h) \\
    \Pi' \mathcal{L}(u') &= - \Pi' (\mathcal{L}u_h-f),
\end{align}
where $\Pi'$ is the $L_2$-projector on the fine-scale basis functions. The Green's function corresponding to the {\color{black}adjoint} of the fine-scale problem is found by solving the following equations
\begin{equation}
    \Pi' \mathcal{L}^*(g'(x,y)) = \Pi' (\delta(x-y) ) \quad \forall x\in \Omega \ \ ; \ \
g'(x,y) = 0 \quad \forall x \in \Gamma.
\end{equation}
The fine-scale can be obtained by super-position as follows:
\begin{equation}
    u'(y) = - \int_{\Omega} g'(x,y)(\mathcal{L}u_h-f)(x)d\Omega_x.
\end{equation}
In the current super-resolution approach, we do not seek $u'$. Rather, we seek  $u'_f$, which is the optimal projection of $u'$ on the finer-space $w_f$. The space of functions in $w_f$ is finer in-comparison to $w_h$ or represents a different kind of space. To this end, the optimal projection of $u'$ on $w_f$ is given by: 
\begin{equation}
    u'_f(y) = \Pi_f u'(y) = - \int_{\Omega} \Pi_f(g'(x,y))(\mathcal{L}u_h-f)(x)d\Omega_x,
\end{equation}
where $\Pi_f(g'(x,y))$ is $L_2$-projection of $g'(x,y)$ on $w_f(y)$. For this case, the fine space can be constructed using higher wavenumber Fourier modes or higher-order Chebyshev polynomials.  
\subsection{Rough Case}
\begin{figure}
    \centering
    \includegraphics[width=0.9\textwidth]{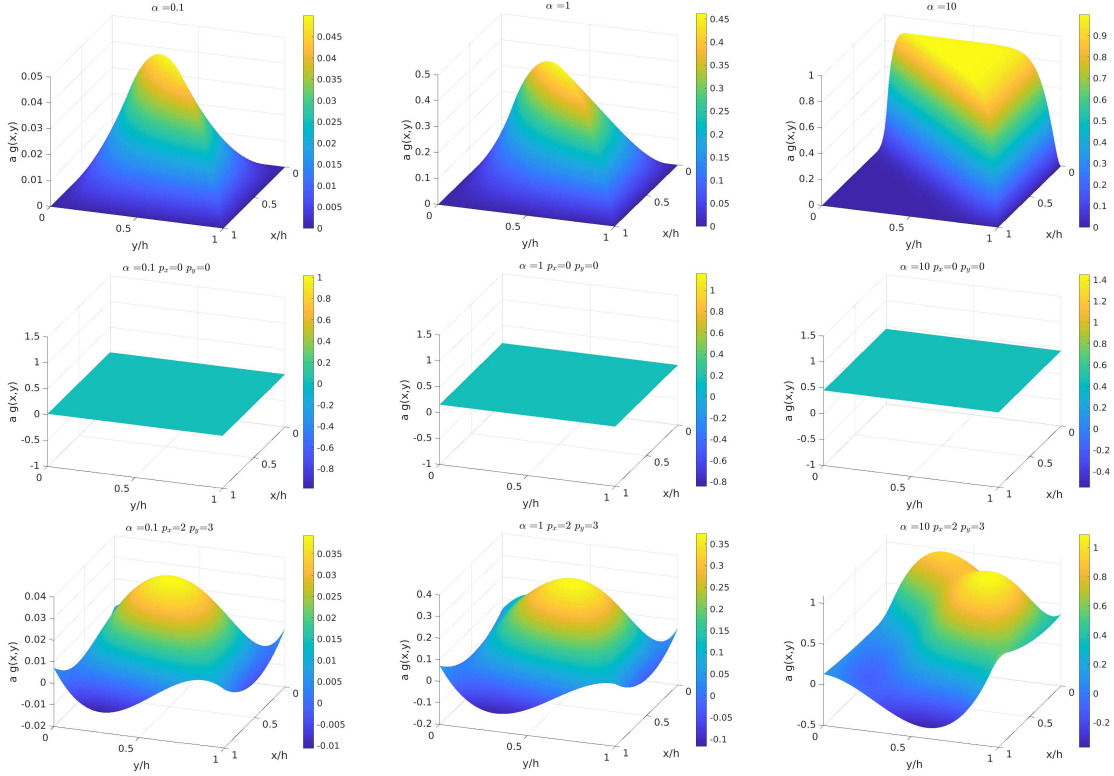}
    \caption{$L_2$ optimal approximation of the fine-scale Green's function on various tensor-product polynomial basis function g' for different cell Peclet number $\alpha$.}
    \label{fig:green}
\end{figure}

\begin{figure}
    \centering
    \includegraphics[width=1.0\textwidth]{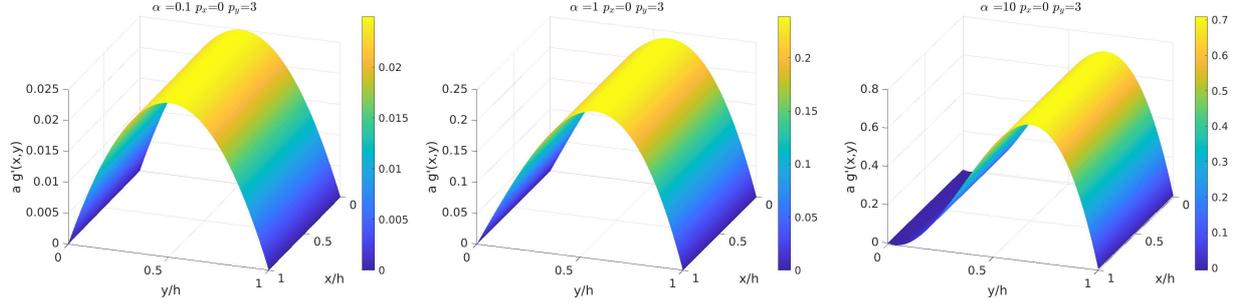}
    \caption{$L_2$-optimal approximation of the fine-scale Green's function on $p_x = 0$ and $p_y = 3$ basis such that $u'_f(y)$ is zero on the boundaries.}
    \label{fig:bubble}
\end{figure}

In the 'rough case', the lack of continuity of derivatives at element interfaces requires us to account for the distributional effects explicitly \cite{VMS}. This case is typical of the finite element method, where piece-wise continuous polynomial functions are used within each element. Similar to the 'smooth case', Hughes et al. \cite{VMS} showed that the exact form of sub-scales in the case of finite elements is given by: 
\begin{equation}
    u'(y) = - \sum_e \left( \int_{\Omega_e} g'(x,y)(\mathcal{L}u_h-f)(x)d\Omega_x + \int_{\Gamma_e}g'(x,y)(bu_h)(x)d\Gamma_x \right). 
    \label{subscale_rc}
\end{equation}
{\color{black}It has to be mentioned that the 'smooth case' can be considered as a limiting case of the 'rough case' when a single element is used and element interfaces are not present.} The sub-scale solution depends on the coarse-scale inside the element and its neighbors. Applying the projection operator on both sides of equation \eqref{subscale_rc} we obtain
\begin{equation}
    u'_f(y) = \Pi_f u'(y) = - \sum_e \left( \int_{\Omega_e} \Pi_f (g'(x,y))(\mathcal{L}u_h-f)(x)d\Omega_x + \int_{\Gamma_e} \Pi_f (g'(x,y))(bu_h)(x)d\Gamma_x \right).
\end{equation}
An approximation to the above equation is given in the form of compact bubble functions which vanish at the element boundaries \cite{BUBBLE,BUBBLE1,BUBBLE2,BUBBLE3}. For 1-D linear problems, solving the element Green's function leads to the coarse-scale solution being the endpoint interpolant of the actual solution \cite{VMS}. Assuming that the coarse-scale is given in the form of the endpoint interpolant of the true solution, the fine-scale solution is given by 
\begin{equation}
    u'_f(y) = \Pi_f u'(y) =  -\int_{\Omega_e} \Pi_f (g'(x,y))(\mathcal{L}u_h-f)(x)d\Omega_x.
    \label{subscaleP}
\end{equation}
A point to note is that application of the projection operator $\Pi_f(y)$  on the element Green's function $g'(x,y)$ leads to the reduction of dimension only in y, i.e.

\begin{equation}
    g'_f(x,y) = \Pi_f g'(x,y) =  \sum_{i} g'_{x,i}(x) \psi_i(y/h),
\end{equation}
where the basis coefficients $g'_{x,i}(x)$ are functions of $x$, which are not necessarily polynomials. However, if the polynomial order $p_c$ of the coarse-scale is given, the coarse-scale residual for (e.g.) the convection-diffusion equation will be of polynomial order $p_C-1$. Thus, projecting $\sum g'_{x,i}(x)$ onto the space of polynomials with order $p_C-1$  and representing $g'(x,y)$ onto tensor-product basis functions in $x$ and $y$ is sufficient. The polynomial order of $y$ is determined by the polynomial order of fine-scales, whereas the polynomial order of x is decided by the maximum polynomial order arising in the coarse-scale residual. For the convection-diffusion problem,  the element Green's function is given by:
\[
    g(x,y)= 
\begin{cases}
    C_1(y)\left(1-e^{-2\alpha x/h}\right),& \text{if } x\leq y\\
    C_2(y)\left(e^{-2\alpha x/h}-e^{-2\alpha}\right),  & x\geq y\\
\end{cases}
\]
where $\alpha$ is the cell Peclet number $ \alpha = {\frac{ah}{2\kappa}}$
{\color{black}and the functions $C_1(y)$ and $C_2(y)$ are defined as
\begin{equation}
    C_{1}(y)=\frac{1-\mathrm{e}^{-2 \alpha(1-(y / h))}}{a\left(1-\mathrm{e}^{-2 \alpha}\right)}, \ \
   C_{2}(y)=\frac{e^{2 \alpha(y / h)}-1}{a\left(1-e^{-2 \alpha}\right)}.
    \label{c2func}
\end{equation}}
The element's Green's function approximated using different order tensor-product basis functions in $x$ and $y$ is shown in Figure \ref{fig:green}. The basis functions used here to approximate the sub-scale  do not necessarily vanish at the element boundary. However, one can also select them to ensure that the sub-scales vanish at the element boundary, similar to a bubble function, as shown in Figure \ref{fig:bubble}. Hence, when the input and output order is fixed, the Green's function for the fine-scales can be represented in a finite number of dimensions. This makes it easier to learn the mapping between the coarse-scale and fine-scale solutions. {\color{black} On further inspection, we find that $ag'(x,y)$ is a function of $\alpha$, $x/h$ and $y/h$. Consequently, $ag_f'(x,y)$ can be written as follows:
\begin{equation}
    ag'_f(x,y)  =  \sum_{i,j} g^{a}_{ij}(\alpha) \phi_i(y/h)\psi_j(x/h),
\end{equation}
where $\phi_i(y/h)$ and $\psi_j(x/h)$ are 1-D basis functions constituting the tensor-product basis functions. Substituting back in equation \eqref{subscaleP}, we obtain:
\begin{equation}
    u'_f(y) =  -\sum_{i} {\phi_i(y/h)} \sum_{j}\int_{\Omega_e} g^{a}_{ij}(\alpha) \psi_j(x/h)  ( \frac{du_h}{dx} - {\frac{\kappa}{a}}\frac{d^2u_h}{dx^2}){d\Omega_x}.
    \label{subscaleP1}
\end{equation}
Next we introduce a constant $u_m$ in the coarse-scale residual calculation as follows: 
\begin{equation}
    u'_f(y) =  -\sum_{i} {\phi_i(y/h)} \sum_{j}\int_{\Omega_e} g^{a}_{ij}(\alpha) \psi_j(x/h)  ( \frac{d(u_h-u_m)}{dx} - {\frac{\kappa}{a}}\frac{d^2(u_h-u_m)}{dx^2}){d\Omega_x}.
    \label{subscaleP2}
\end{equation}
It can be observed that subtracting $u_m$ from $u_c$ does not introduce any error in equation \eqref{subscaleP2} because we are taking derivatives of a constant. Writing the coarse-scale solution and the constant $u_m$ in terms of the nodal Lagrange basis functions $u_h = \sum_k u_{h,k} w_{h,k}(x/h)$ and  $u_m = \sum_k u_m w_{h,k}(x/h)$, and substituting in equation \eqref{subscaleP2} we get
\begin{equation}
    u'_f(y) =  -\sum_{i} {\phi_i(y/h)} \sum_{j}\int_{\Omega_e} g^{a}_{ij}(\alpha) \psi_j(x/h) \sum_k (u_{h,k}-u_m)(w'_{h,k}(x/h) - {\frac{1}{\alpha}} w''_{h,k}(x/h))\frac{d\Omega_x}{h}.
    \label{subscaleP3}
\end{equation}
Dividing both sides with a normalizing parameter $u_{rms}$ (which will be defined later) and re-arranging we obtain:
\begin{equation}
    \frac{u'_f(y)}{u_{rms}} =  -\sum_{i} {\phi_i(y/h)} \sum_k \frac{u_{h,k}-u_m}{u_{rms}} \int_{\Omega_e} \left(w'_{h,k}(x/h) - {\frac{1}{\alpha}} w''_{h,k}(x/h)\right) \sum_{j} \left(g^{a}_{ij}(\alpha) \psi_j(x/h)\right)\frac{d\Omega_x}{h},
    \label{subscaleP4}
\end{equation}
 where the integral $\int_{\Omega_e} \left(w'_{h,k}(x/h) - {\frac{1}{\alpha}} w''_{h,k}(x/h)\right) \sum_{j} \left(g^{a}_{ij}(\alpha) \psi_j(x/h)\right)\frac{d\Omega_x}{h}$ is a function of $\alpha$ which can be written as linear combinations of $g^{a}_{ij}(\alpha)$ and $g^{a}_{ij}(\alpha) \over \alpha$. 
 
 {\bf With this normalization, the problem of learning the sub-scales reduces to learning the projected element Green's function basis coefficients $g^{a}_{ij}(\alpha)$ which define the shapes of the surfaces plotted in Figure \ref{fig:green}.} Equation \eqref{subscaleP4} also suggests that the normalised sub-scale basis coefficients can be written as sum of the products of normalized coarse scales basis coefficients and functions of $\alpha$. Further, the number of such functions of $\alpha$ needed to be learnt are finite. Equation \eqref{subscaleP4} also suggests for a linear problem that the sub-scales depend linearly on the coarse-scale basis coefficients. The dependence on $\alpha$, however, can be non-linear. These insights will be used later in sections 3 and 4. 
}

\section{VMS-inspired feature selection}
\begin{figure}
    \centering
    \includegraphics[width=0.5\textwidth]{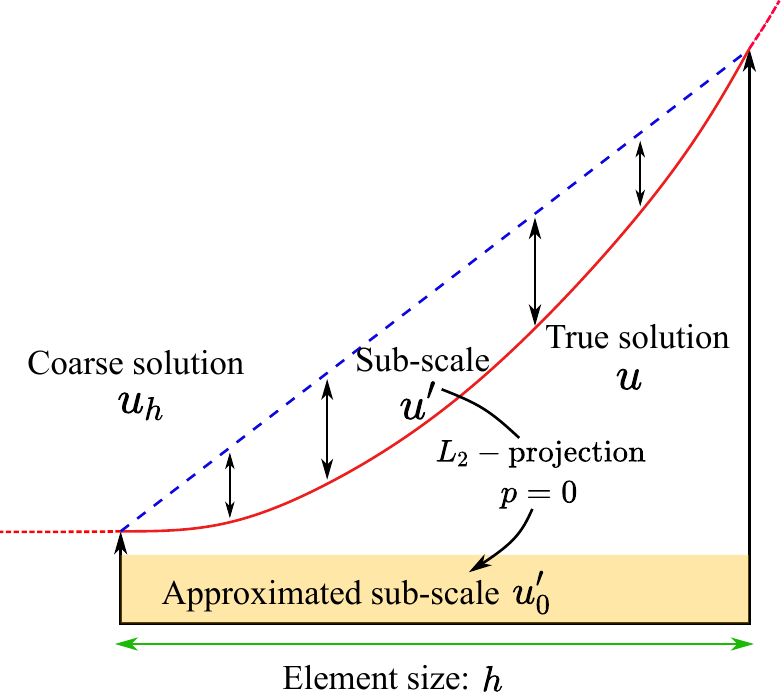}
    \caption{Computation of $u'_0$ by element-wise $L_2$-projection of the sub-scale $u'$ on the $p=0$ polynomial space.}
    \label{fig:CGdata}
\end{figure}

\begin{figure}
    \centering
    \includegraphics[width=0.5\textwidth]{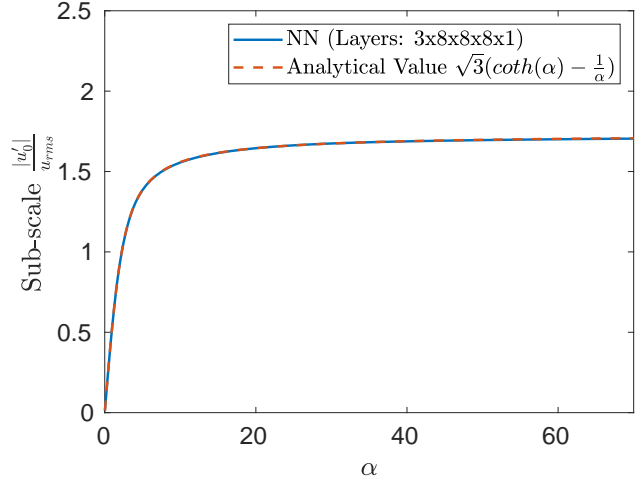}
    \caption{Comparison of sub-scales magnitude as a function of cell Peclet number $\alpha$ obtained analytically vs. that learnt from data using a N-N.}
    \label{fig:tau0}
\end{figure}

In this section, we will derive an appropriate feature set and the network architecture for our model. To demonstrate the action of the super-resolution operator, we assume the coarse space to be composed of piece-wise linear polynomials and the governing PDE to be the linear convection-diffusion equation given by the following differential operator and boundary condition  
\begin{equation}
    \mathcal{L} \triangleq a \frac{d}{dx} - {\kappa}\frac{d^2}{dx^2} \quad in \quad \Omega=[0,L] , \ \ 
    u = 0 \quad on \quad \Gamma=\{0,L\}.
\end{equation}
 For a linear 1-D element of size $h$ existing between $0 \leq x \leq h$ in its local co-ordinates, the coarse solution $u_h$ in terms of the end-point values $u(0)$ and $u(h)$ is given by:
\begin{equation}
    u_h(x) = u(0)(1-x/h) + u(h)(x/h).
\end{equation}
 The space in which the fine scales are approximated can be a discontinuous finite element space or a bubble function space. The $L_2$-optimal approximation of the fine scales $u'$ on any polynomial space existing inside an element is given by the projection $\Pi_f$ :
\begin{equation}
    u'_f(y) = \Pi_f u'(y) =  -\int_{\Omega_e} \Pi_f (g'(x,y))a\left({u(h)-u(0)\over h}\right)d\Omega_x.
    \label{pgreen}
\end{equation}
The coarse solution considered here is an endpoint interpolant of the true solution. In that case, one can determine the exact form of the sub-scale inside the element by assuming the endpoint values as Dirichlet boundary conditions and by solving the equation inside the element:
\begin{equation}
    u'(y) = (u(h)-u(0))\left({1-e^{2\alpha {y\over h}}\over 1-e^{2\alpha }}-{y\over h}\right).
\label{1D_exact}
\end{equation}
The simplest approximation of the bubble function is obtained by projecting it on a $p_0$ discontinuous space inside the element i.e. 
\begin{equation}
    u'_{0}(y) =\int_{0}^{h} (u(h)-u(0))\left({1-e^{2\alpha  {y\over h}}\over 1-e^{2\alpha }}-{y\over h}\right) dy = -{a{{(u(h)-u(0))}\over h}} {h \over {2a}}\left(coth (\alpha) - {1\over \alpha}\right),
    \label{famoustau}
\end{equation}
where the first part is the residual when linear basis functions are used, and the second part is the form of the stabilization parameter $\tau$ commonly used in stabilized methods. {\color{black} For the linear CG finite element method, Equation \eqref{famoustau} represents the closure for obtaining the nodally exact solution\cite{VMS}. Equation \eqref{famoustau} can also be obtained by first projecting the elements Green's function on $p=0$ discontinuous space i.e.
\begin{equation}
    g'_{f,0}(x,y) = \frac{1}{h^2} \int_{\Omega_e} \int_{\Omega_e} g'(x,y) d\Omega_x d\Omega_y = {1 \over {2a}}\left(coth (\alpha) - {1\over \alpha}\right),
    \label{greens project}
\end{equation}
and using this result in equation \eqref{pgreen} to evaluate the sub-scale as follows:
\begin{equation}
    u'_0(y) = \Pi_{f,0} u'(y) =  -\int_{\Omega_e} {1 \over {2a}}\left(coth (\alpha) - {1\over \alpha}\right){a{{(u(h)-u(0))}\over h}}d\Omega_x = -{a{{(u(h)-u(0))}\over h}} {h \over {2a}}\left(coth (\alpha) - {1\over \alpha}\right).
\end{equation}
}
Next, we define the mean and r.m.s quantities of the coarse solution in an element as follows:  
\begin{equation}
    u_m = {\int_0^h u_h d\Omega \over h}  \ \ ; \ \
    u_{rms} = \sqrt{\int_0^h (u_h-u_m)^2 d\Omega \over h}.
\end{equation}
An important observation is that the solution is independent of the mean $u_m$:
\begin{equation}
    u'(y) = ((u(h)-u_m)-(u(0)-u_m))\left({1-e^{2\alpha {y\over h}}\over 1-e^{2\alpha }}-{y\over h}\right).
\end{equation}
If our approximation space is linear, then $u_{m}$ and $u_{rms}$ are given by: 
\begin{equation}
    u_{m,1} \triangleq (u(0)+u(h))/2 \ \; \ \
    u_{rms,1} \triangleq {|u(h)-u(0)|\over \sqrt{12}}.
\end{equation}
Re-arranging this form we get:
\begin{equation}
    {u'_{0}(y) \over {u_{rms,1}}} = \sqrt{3}({1\over \alpha} - coth (\alpha))sgn(u(h)-u(0)),
    \label{idea_arch}
\end{equation}
{\color{black}where $sgn$ denotes the sign function}. The above equation can be simplified as follows:
\begin{equation}
    {u'_{0}(y) \over {u_{rms,1}}} = 
\begin{cases}
    \sqrt{3}({1\over \alpha} - coth (\alpha)),  & {u(h)-u(0) \over |u(h)-u(0)|}\geq 0\\
   -\sqrt{3}({1\over \alpha} - coth (\alpha)).  & {u(h)-u(0) \over |u(h)-u(0)|}\leq 0\\
\end{cases}
\label{para_dep}
\end{equation}
If we compute the the mean-subtracted basis-coefficients of the coarse solution $u_h$ and re-scale them with the r.m.s $u_{rms,1}$ we get:
\begin{equation}
    {u(0)-u_m \over u_{rms,1}} = \sqrt{3}{u(0)-u(h) \over |u(h)-u(0)|}, \ \ ; \ \
    {u(h)-u_m \over u_{rms,1}} = \sqrt{3}{u(h)-u(0) \over |u(h)-u(0)|}.
\end{equation}
These parameters determine the sign of the sub-scale in equation \eqref{para_dep}. Hence, the magnitude of the sub-scales is fully determined by the physics-informed parameter $\alpha$ and its sign (phase) is determined by these parameters. Consequently, an appropriate choice of the feature will be:
\begin{equation}
    {u'_{0}(y) \over {u_{rms,1}}} = f\left(\alpha,{u(0)-u_m \over u_{rms}},{u(h)-u_m \over u_{rms}}\right).
\end{equation}
The two extra parameters, in this case, are redundant because they are the same in magnitude and opposite in sign. Hence, only one parameter can be used:
\begin{equation}
    {u'_{0}(y) \over {u_{rms,1}}} = f\left(\alpha,{u(h)-u(0) \over u_{rms}}\right).
    \label{eq:functional_form}
\end{equation}
{\color{black}
  Although equations \eqref{famoustau} and \eqref{eq:functional_form} are identical, the generalisibilty of a neural-network model is considerably affected by the choice of the feature set and the normalization process. For example, the model form 
 \begin{equation}
    u'_{0}(y) = g\left(a,\kappa,h,u(h),u(0)\right),
 \label{badmodel}
 \end{equation}
 does not utilize the idea that only specific combinations of $a,\kappa,h$ i.e. cell Peclet number $\alpha$ effect the sub-scale solution distinctively. To train this model, a big data-set with a large stretch in the values of the input parameters is required. This equation is also not invariant to the units or the scaling used for the input parameters. The proposed model form in equation \eqref{eq:functional_form} tries to address some of these challenges. An important point to note is that the functional form presented in equation \eqref{eq:functional_form} is derived for the advection-diffusion problem. However, this structure will serve as the inspiration for applying this method to other non-linear problems, as detailed in the following section. } 
{\color{black}
\section{Learning VMS-consistent subscales}
In this section, we will attempt to derive a general model structure that can be used to learn sub-scales arising in a wide range of multi-scale problems. In case of super-resolution, the sub-scales contain the fine-scale information that is absent in the lower-resolution image. When used as the closure, they are responsible for modeling the effect of unresolved fine-scales on the coarse-scales. Irrespective of the mode of application, the model structure should not be different.} As a first step, we learn the mapping presented in equation \eqref{eq:functional_form}  and compare it to the analytical solution. Data is first obtained by solving the equation at different Peclet numbers $Pe$ on very fine meshes (we refer to this as the DNS) for training the network. Similarly, we can also generate data by dividing a single high $Pe$ DNS case into multiple cases with different element sizes, i.e., different cell Peclet numbers. {\color{black}The coarse-scale is obtained in the form of end-point interpolant, and the approximation to the sub-scale is then computed for each such element numerically as shown in figure \ref{fig:CGdata}}. A small network with a size of $3\times8\times8\times8\times1$ is then trained using this data. Figure \ref{fig:tau0} shows the comparison between the sub-scale obtained analytically vs. that learned purely from data using neural network (N-N). It can be observed that the small network could learn the analytical solution accurately. {\color{black} The discussion in this section was mainly focused on learning the sub-scales. In appendix A, we demonstrate how these sub-scales can be used as closures for the CG finite element method and further extend them to high order discretizations.}

One way of obtaining the exact form of the sub-scale is by keeping our approximating space of the coarse-scale the same (linear) and increasing the order of the discontinuous space $p$ in which $u'$ is approximated. In limit, $p \to \infty$ the approximate sub-scale should approach $u'(y)$. However, the order of the polynomial $p$ required to learn the function increases when $Pe$ is increased. By limiting the output order $p$ to the order of super-resolution, we are reducing the complexity and size of the network, as discussed in the last section. This is because, when $\alpha \to \infty$, $f(\alpha) = {|u'|\over u_{rms}} = \sqrt{3}(coth(\alpha)-{1\over \alpha})$ is well-behaved, whereas if $u'(y)$ given by the equation \eqref{1D_exact} is learnt directly as a function of $y$ and $\alpha$, the function becomes steep at $y=h$ for $\alpha \to \infty$. A solution to this problem is to use features such as  ${1-e^{2\alpha {y\over h}}\over 1-e^{2\alpha }}$ as inputs. However, this restricts the method to one kind of problem. Similarly, the optimal form of the sub-scale on discontinuous $p=1$ basis functions is given by:
\begin{equation}
    {u'_1(y)  \over {u_{rms,1}}} = C'_1(\alpha)\psi_1(y/h) + C'_1(\alpha)\psi_2(y/h).
\end{equation}
The form of the function to be learned for this case is:
\begin{equation}
    \left[C'_1,C'_2\right] = \mathbf{f}(\alpha).
\end{equation}
The above analysis was performed for the continuous Galerkin (CG) method, but these ideas can be extended to the discontinuous Galerkin (DG). Retaining the same structure, we extend the technique to non-linear/linear problems for both CG/DG types of basis functions as follows:
\begin{align}
    [C'_{1,p_s},C'_{2,p_s},..,C'_{(p_s+1)^d,p_s}] = \mathbf{f} \biggr(\alpha,\left[\tilde{C}_{1,p_c},\tilde{C}_{2,p_c},..,\tilde{C}_{(p_c+1)^{d},p_c}\right],..   \label{fform} \\ 
    ,\left[\tilde{C}_{1,p_c},\tilde{C}_{2,p_c},..,\tilde{C}_{(p_c+1)^d,p_c}\right]_m,..\biggr), \nonumber
\end{align}
where $p_s$ and $p_c$ are the polynomial orders of the spaces in which the sub-scale and coarse-scale are optimally represented, and $d$ denotes the dimension of the problem. This function, apart from $\alpha$ (equivalent to cell $Re$/ cell $Pe$) also contains the basis coefficients of the current element and its neighbors. The term $\left[\tilde{C}_{1,p_c},\tilde{C}_{2,p_c},..,\tilde{C}_{(p_c+1)^d,p_c}\right]_m$ with sub-script $m$ denotes the mean subtracted normalized basis coefficient of $m^{th}$ neighbour. The neighbour information is critical when discontinuous basis is used, or when bubble function approximation are not employed in CG, or non-local transfer of information happens from outside the element. These coefficients are first subtracted with the coarse scale mean $u_m$ and then normalised with the coarse scale R.M.S. $u_{rms}$ as done previously. The output of the function $\left[C'_{1,p_s},C'_{2,p_s},..,C'_{(p_s+1)^d,p_s}\right]$ denotes the basis coefficients of the sub-scale that has been normalised with $u_{rms}$ only i.e.
\begin{align}
   \left[\tilde{C}_{1,p_c},\tilde{C}_{2,p_c},..,\tilde{C}_{(p_c+1)^d,p_c}\right]_m = {\left[{C}_{1,p_c}-u_m,{C}_{2,p_c}-u_m,..,{C}_{(p_c+1)^d,p_c}-u_m\right]_m/{u_{rms}}},
   \label{basic_frm}
\end{align}
where $C_{i,j}$ denotes the actual basis coefficients. The quantities used for shifting and non-dimensionalizing the input parameters,i.e., $u_m$ and $u_{rms}$ respectively, and non-dimensionalizing the output parameters $u_{rms}$ are calculated using the coarse-scale solution in the center element only. As will be seen later in this paper, the non-dimensionalization process is critical for the N-N model to generalize. The output coefficients are finally re-scaled with $u_{rms}$ and added to the coarse-scale to obtain the super-resolved solution as follows:
\begin{equation}
   u_{sr} = u_{p_c} + u'_{p_s} = u_{p_c} + u_{rms}\sum_i^{(p_s+1)^d} C'_{i,p_s}\psi_{i,p_s},
\end{equation}
where, $\psi_{i,p_s}$ denotes basis function corresponding to the $i^{th}$ node or mode. Division by 
${u_{rms}}$ in equation \eqref{basic_frm} is a problem when ${u_{rms}}$ is precisely equal to zero. However, adding a small positive number $\epsilon$ to ${u_{rms}}$ while dividing was sufficient for all the cases presented below. 

\subsection{The Variational Super-resolution Network architecture}
\begin{figure}
\centering
\includegraphics[width=0.7\textwidth]{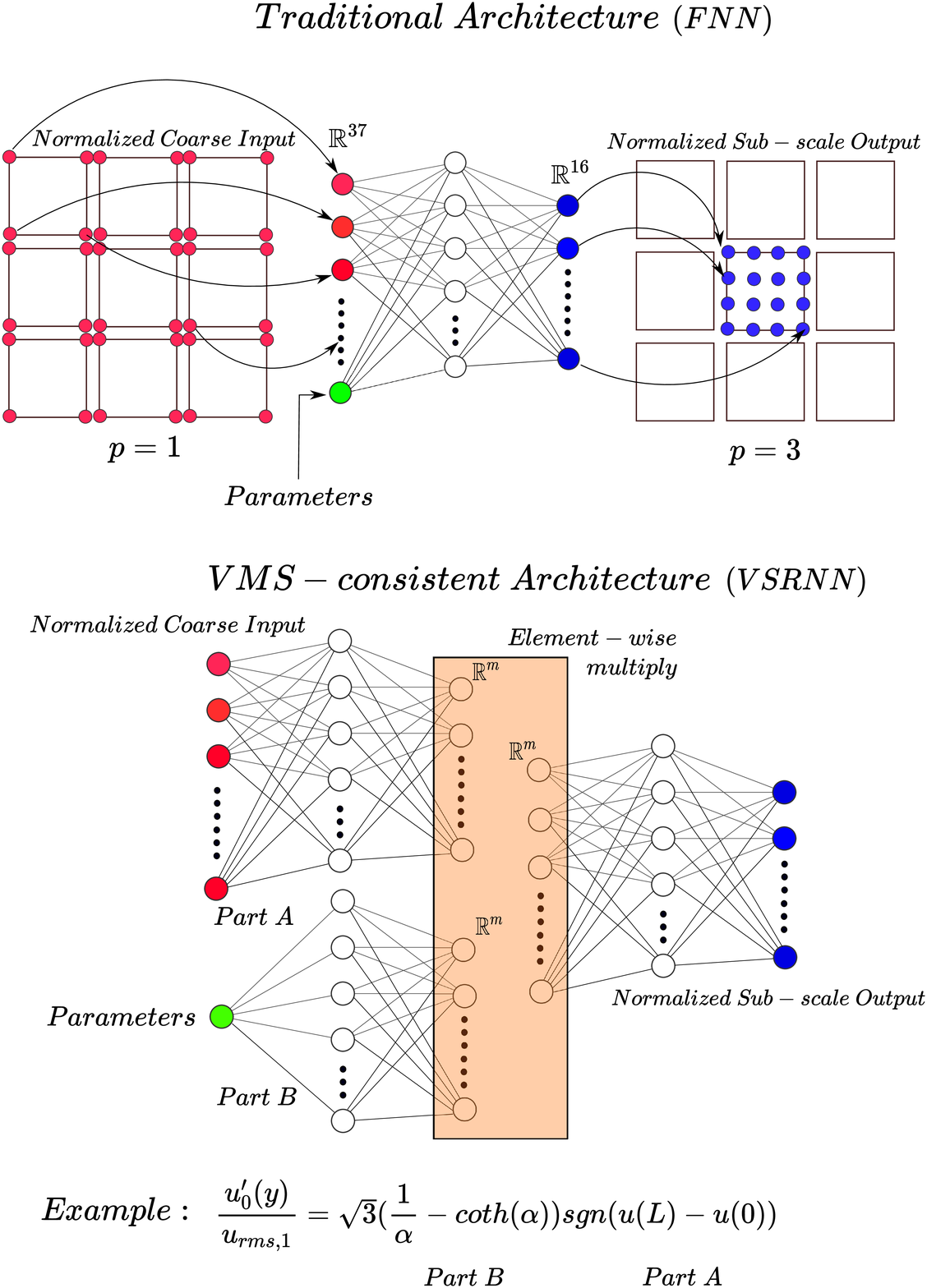}
\caption{VMS-consistent architecture and features are used for learning the mapping in VSRNN.}
\label{archi}
\end{figure}

In addition to the model features, the model architecture can be made consistent with the VMS formulation. As proposed in equation \eqref{fform}, the input to the model are the physics-informed parameters such as the cell Péclet number $\alpha$, along with the normalized mean-subtracted coarse-scale basis coefficients of an element and its neighbor. The output to the network are the normalized sub-scale basis coefficients in that particular element. The physics-informed parameter can also be other non-dimensional numbers such as the CFL number or the cell Reynolds number $Re_{\Delta}$ specific to the problem. Given these sets of input and output features, many possible ways of embedding them into the model exist. Figure \ref{archi} shows two different kinds of network architectures to achieve this. 

The traditional approach is based on the idea of training one single fully connected N-N with both the normalized coarse-scale basis coefficients and the physics informed-parameter as inputs. If the normalized sub-scale basis coefficients are denoted by $\mathbf{u}'$, the normalized input coarse basis coefficients of the element and its neighbors as $\mathbf{u}_c$ and the physics-informed parameter $\alpha$, then the traditional model is given by:

\begin{equation}
  \mathbf{u}'  = \mathbf{f}^{FNN}\left(\alpha,\mathbf{u}_c\right),  
\end{equation}
where $\mathbf{f}^{FNN}$ denotes a fully-connected neural network (FNN). Another approach also called the variational super-resolution N-N (VSRNN), is based on a multiplicative strategy in which the fine-scales are approximated by a sum of products of individual functions of the parameters and the coarse-scales. The model form can be summarized as follows:
\begin{equation}
  \mathbf{u}'  = \mathbf{f}^{FNN}(\mathbf{g}_{\alpha}\odot \mathbf{g}_{u}),  \ \ ; \ \
  \mathbf{g}_{\alpha}  = \mathbf{h}^{FNN}(\alpha),  \ \ ; \ \ 
  \mathbf{g}_{u}  = \mathbf{k}^{FNN}(\mathbf{u}_c),  
\end{equation}
where $\mathbf{f}^{FNN}$, $\mathbf{h}^{FNN}$ and $\mathbf{k}^{FNN}$ denote three different FNNs. The symbol $\odot$ denotes element-wise multiplication between two vectors of the same size. {\color{black}This architecture is inspired by equation \eqref{subscaleP4} and the analytical solution of the sub-scale provided in equations \eqref{idea_arch}}. In this case, $\mathbf{g}_{\alpha}$ (Part B) learns the dependence of $\alpha$ i.e. $\sqrt{3}\left({1\over \alpha} - coth (\alpha)\right)$ and $\mathbf{g}_{u}$ (Part A) learns the dependence of the normalized coarse-scale basis coefficients, i.e., $sgn(u(h)-u(0))$.  {\color{black} In sections 2-4 and appendix A, we tried to develop insights into the working of the VSRNN and demonstrated its application as a closures for the CG method. In sections 5-7, we use the VSRNN to perform super-resolution and sub-grid modelling for the DG method. This does not imply that the exact sub-scales used for the CG method in sections 2-4 are re-used in sections 5-7. Only the model form has been assumed to be the same which is finally trained on the correct sub-scale that is consistent with the DG approach. Details about the scale-decomposition used for the DG method are detailed in section 5.}

\section{Data Generation}
The generation of proper training and testing data is as critical as the model architecture and features.  For example, low-resolution data can be obtained from a variety of high fidelity sources (simulations and experiment) and can be coarse-grained. There is no guarantee that a model trained to perform super-resolution of the filtered solution will be useful unless the filtering operation is consistent with the underlying numerics. To this end, consider the VMS decomposition of the full-order solution $u$ as follows:
\begin{equation}
u = {u_h} + u',
\label{vmsuw}
\end{equation}
where ${u_h} \in {\mathcal{V}_h}$ and $u' \in \mathcal{V}'$. The vector space of functions $\mathcal{V}\equiv\mathcal{H}^1(\Omega)$ is a Sobolev space where the functions and their derivative are square-integrable. This space is now decomposed as follows:
\begin{equation}
\mathcal{V} = {\mathcal{V}_h} \oplus \mathcal{V}',
\end{equation}
where $\oplus$ represents a direct sum of ${\mathcal{V}_h}$ and $\mathcal{V}'$. Let us also define $\mathcal{T}_{h}$ to be a tessellation of domain $\Omega$ into a set of non-overlapping elements, $K$, each having a sub-domain $\Omega_k$ and boundary $\Gamma_k$.  ${\mathcal{V}_h}$ is now defined as:
\begin{equation}
{\mathcal{V}_h} \triangleq \left\{u\in{L_2}(\Omega):u|_{\mathcal{T}}\in P^k(T),T\in\mathcal{T}_{h} \right\},
\end{equation}
where the space of polynomials up to degree $k$ is denoted as $P^k$. Defining ${\mathcal{V}_h}$ in this manner allows discontinuity in the solution across element boundaries. Given $u$ from the high-fidelity simulation, our goal is to find the optimal representation of $u$ in the coarse sub-space $\tilde{\mathcal{V}}$. In our case, we will use the $L^2$ projection to obtain $u_h$ which minimises the value of $||u-u_h||_2^2$. This problem is equivalent to the problem of finding $u_h\in \tilde{\mathcal{V}}$ such that   
\begin{equation}
(u,w_h)=(u_h,w_h)  \quad \forall {\tilde{w}} \in {\tilde{\mathcal{V}}}.
\label{proj}
\end{equation}

\begin{figure}[h]
\centering
\includegraphics[width=0.60\textwidth]{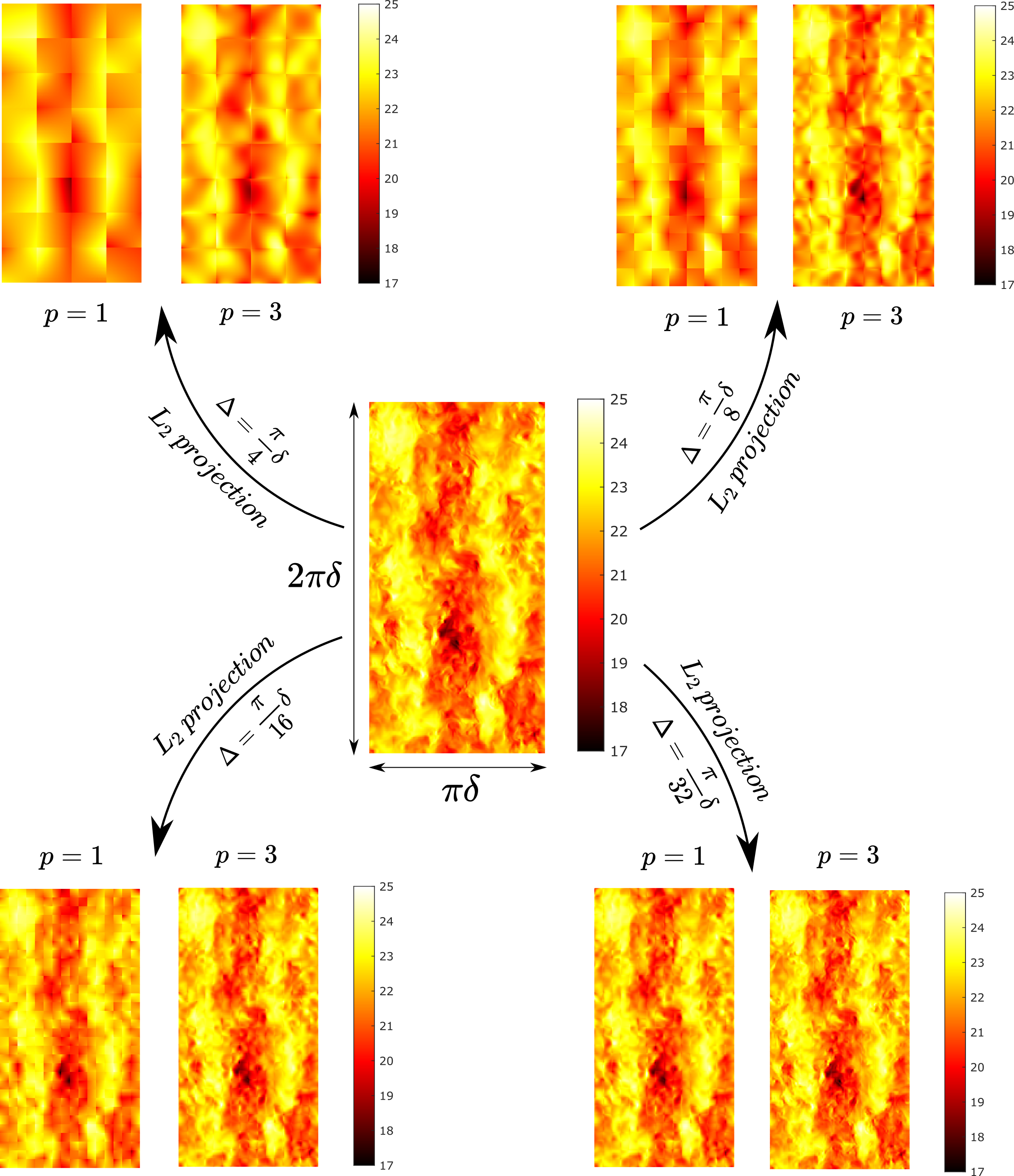}
\caption{Example $L_2$-projected snapshots of the DNS data for channel flow at $Re_\tau \approx 950$ and wall normal distance of $y^+ \approx 910$.}
\label{l2proj}
\end{figure}

For example, to generate training data for section 6, we use direct simulation (DNS) results for a channel flow at $Re_{\tau}\approx950$ \cite{lozano2014effect}. The 3-D data is sliced into many 2-D planes at different $y^+$ locations, and projection is performed in 2-D for simplicity. The fine space and coarse space's polynomial orders are chosen to be 3 and 1, respectively, i.e., we are super-resolving $p=1$ results to $p=3$ as shown in figure \ref{l2proj}. 

The computation of terms on the RHS of the equation \eqref{proj} requires special care. Although $u$ has been assumed to exist in an infinitely high dimensional space, in reality, it is not. For example, the Kolmogorov scale $\eta$ dictates the size of the smallest size eddy and the size of $u$. {\color{black}Although the dimension of $u$ is finite, it is enormous when compared to $u_h$ because the size of our finite element grid $h$ is much greater than $\eta$}. Hence, to accurately compute these terms, the DNS data is first interpolated using cubic-splines to a much finer-grid $O(\eta)$ and then the inner products with the coarse finite element basis function (having dimension $O(h)$) are computed on these fine-meshes using the Simpson's Rule.  {\color{black}Interpolation of DNS was done to ensure that the projected solution $u_{h}$ did not change significantly due to the numerical integration scheme}.
Sample $L_2$-projected snapshots of the DNS data on elements of different sizes and orders are shown in figure \ref{l2proj}.   

\section{Application to Linear Advection}\label{sec:linadv}
In this section, we apply our super-resolution methodology to the linear advection equation in the domain $\Omega \subset \mathbb{R}$ with the boundary $\Gamma = \partial \Omega$ as follows
\begin{equation}
{\partial u \over \partial t} + a{\partial u \over \partial x} =  0,
\label{adv}
\end{equation}
with time $t \in ( 0,T \rbrack$, and spatially periodic boundary conditions on $\Gamma$. The parameter $a$ denotes the advection velocity.  The required training data is generated by $L_2$-projecting the true solution on coarse and fine spaces. Unlike traditional methods, in which only the spatial term in the PDE is discretized using finite elements, we will consider 2-D space-time finite elements spanned by $p=1,2$ degree tensor-product Lagrange basis functions in space and time. The goal is to investigate the application of our super-resolution method in two different settings. First, as a model to super-resolve coarse low-order finite element data to high-order finer finite element data. Second, as a method to improve the existing finite element method for this problem in a predictive setting on a problem with a very different initial condition in comparison to the training data.

\begin{figure}[h]
\centering
\includegraphics[width=0.8\textwidth]{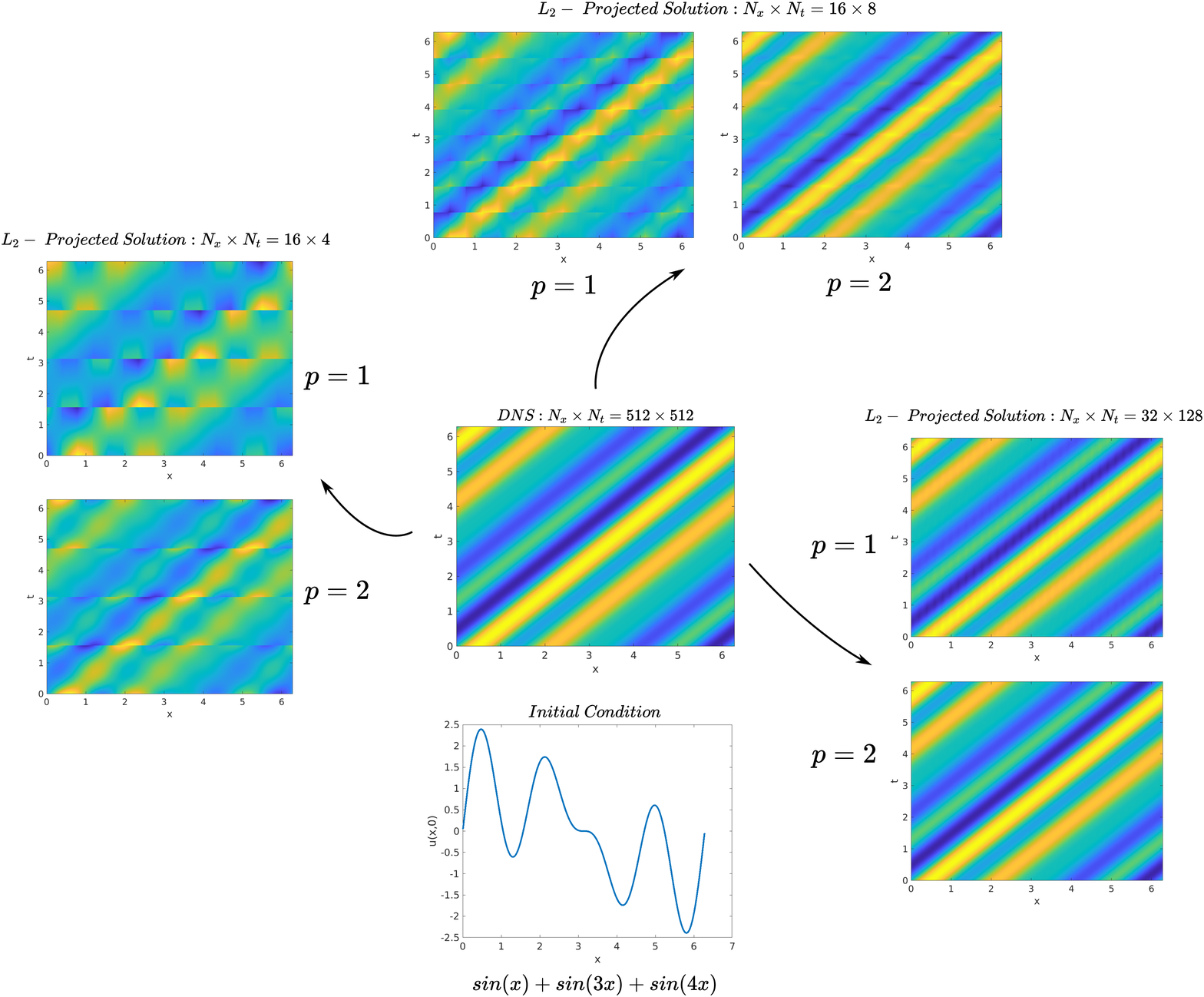}
\caption{High-resolution linear advection solution on a 512x512 mesh is $L_2$-projected on different finite element meshes for $p=1,2$ to generate data for training the model.}
\label{l2proj_ad}
\end{figure}

\begin{figure}[h]
\centering
\includegraphics[width=0.9\textwidth]{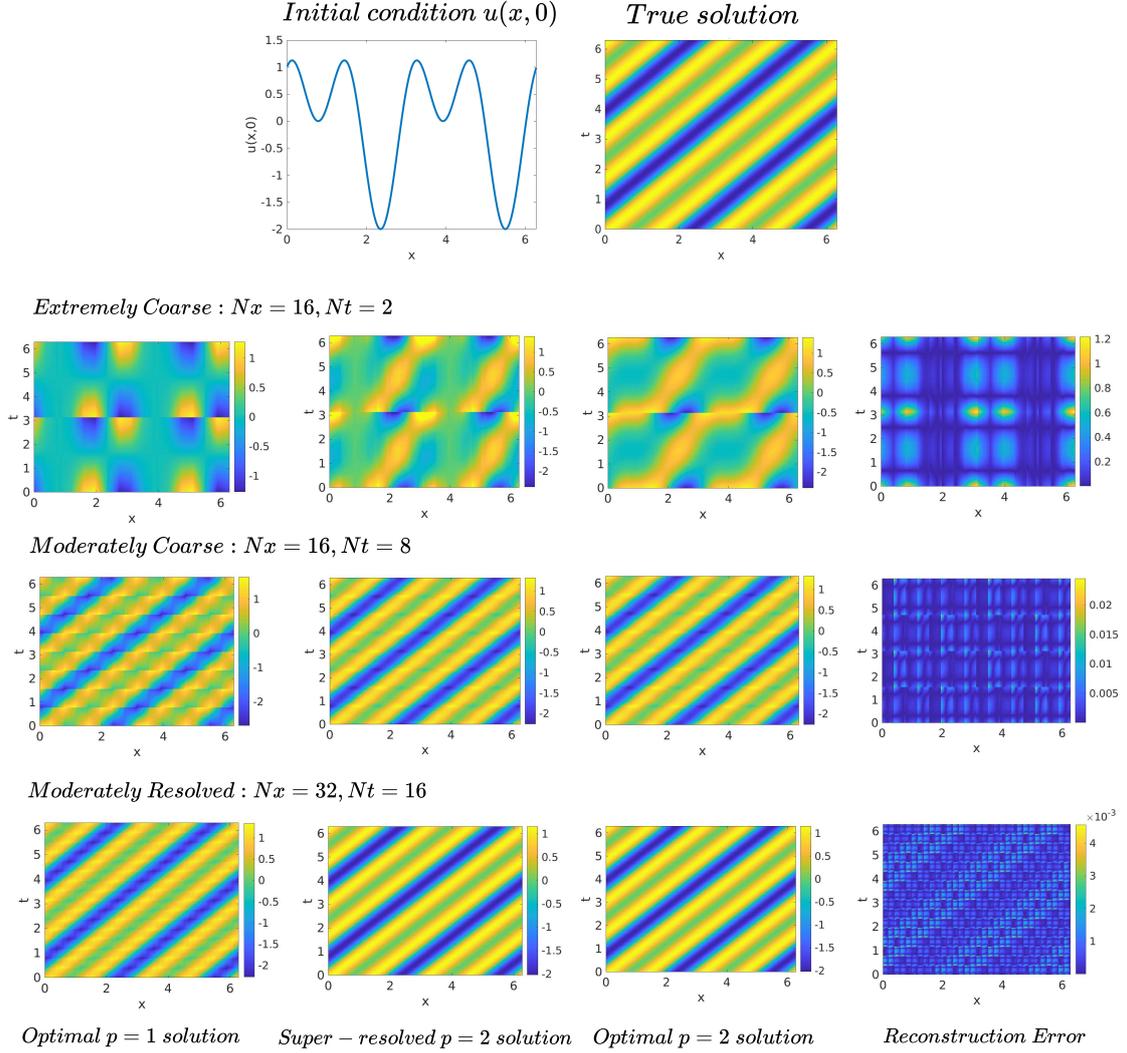}
\caption{Comparison of the true $L_2$-projected $p=2$ solution and that obtained by super-resolution of $L_2$-projected $p=1$ solution on the space-time plane.}
\label{ad_offline_12}
\end{figure}

\begin{figure}[h]
\centering
\includegraphics[width=0.6\textwidth]{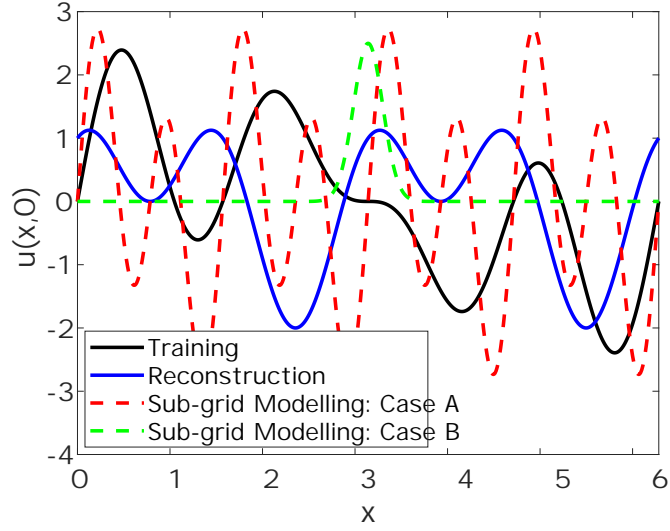}
\caption{Different initial conditions are used for training, offline reconstruction and online evaluation of the model.}
\label{init_cond}
\end{figure}

\begin{figure}[h]
\centering
\includegraphics[width=1.0\textwidth]{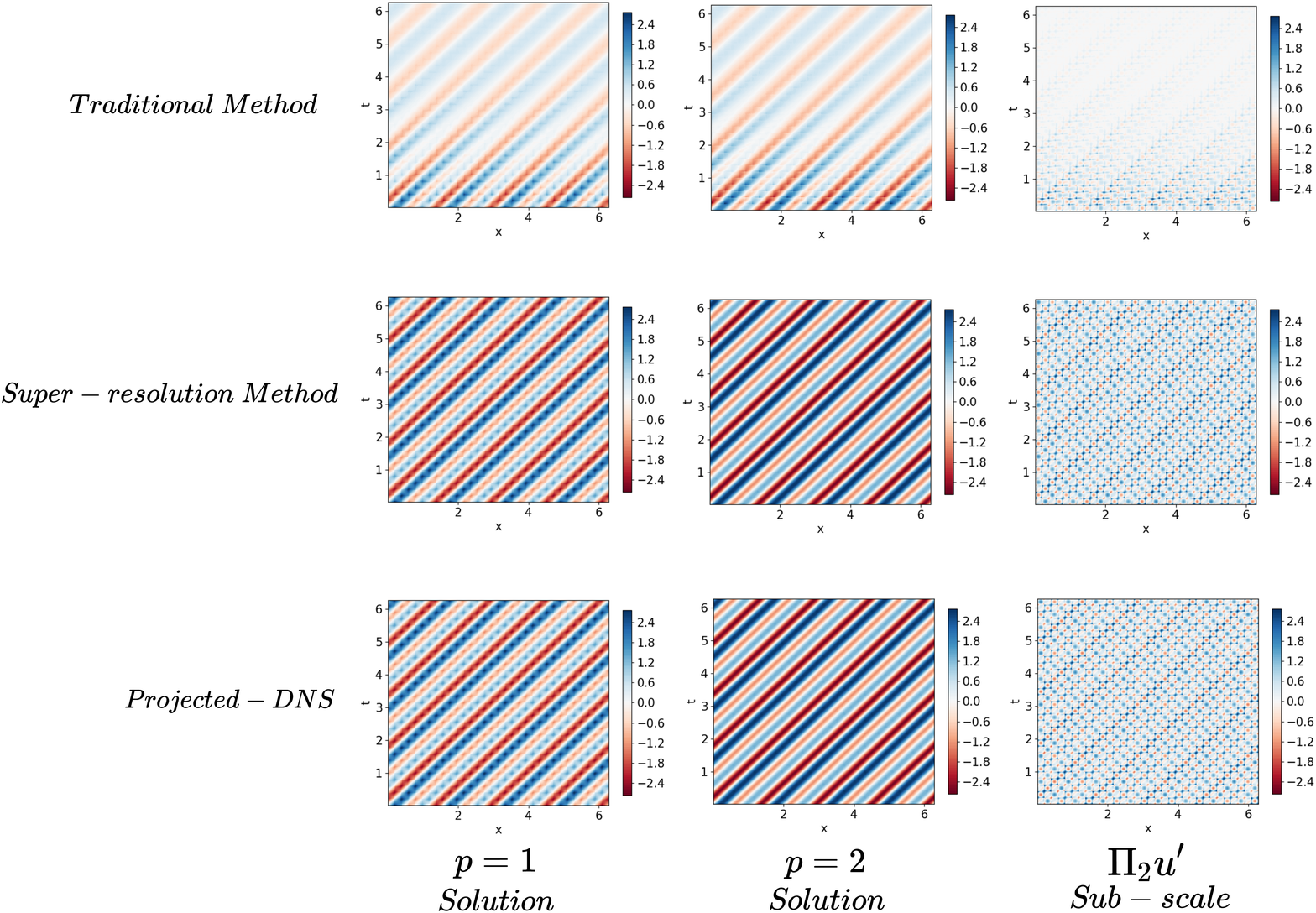}
\caption{Comparison of solution obtained using the traditional space-time method and the super-resolved method to the projected DNS solution for initial condition $u(x,0)=sin(4x)+2sin(8x)$.}
\label{AD_online}
\end{figure}

\begin{figure}[h]
\centering
\includegraphics[width=1.0\textwidth]{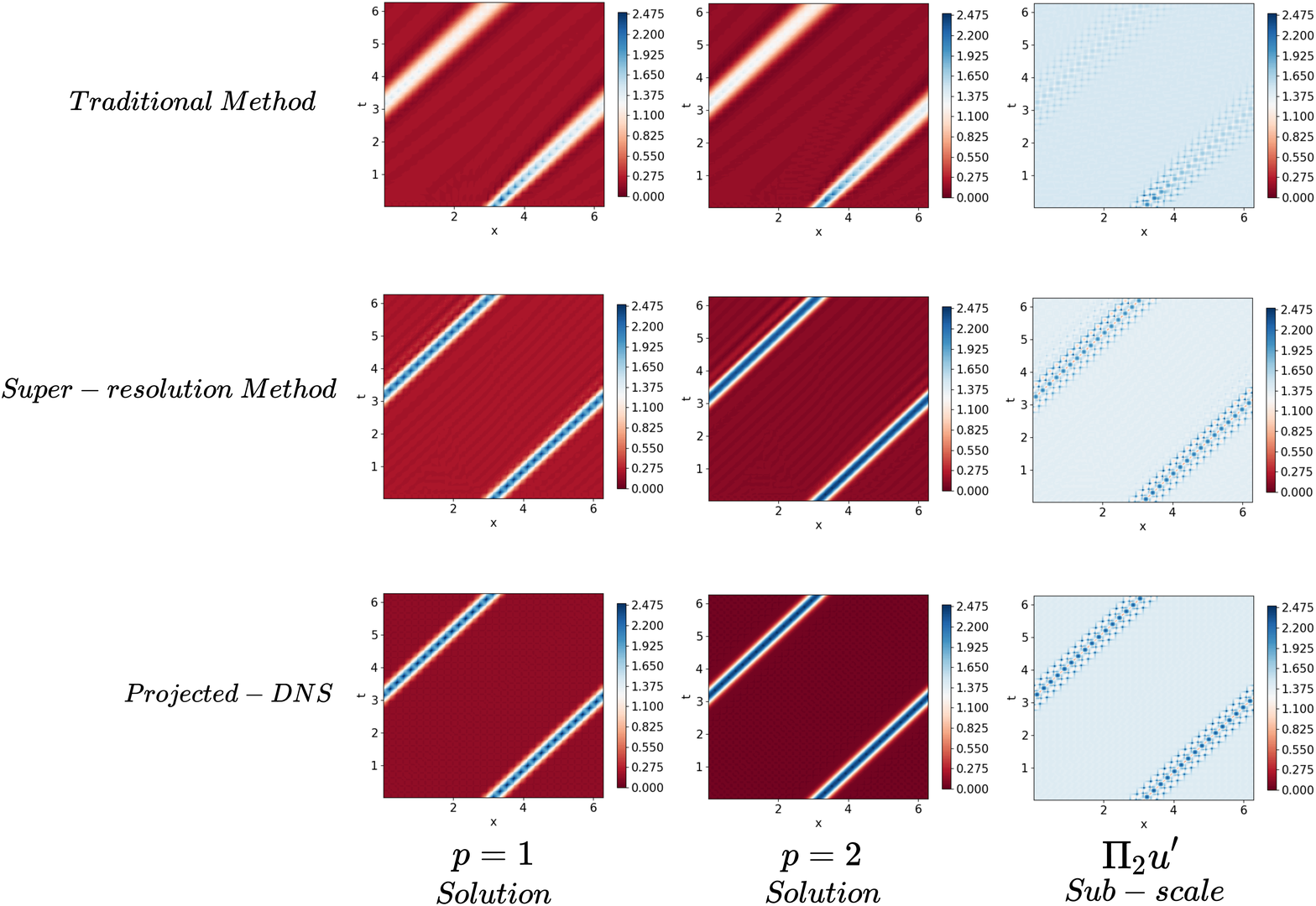}
\caption{Comparison of solution obtained using the traditional space-time method and the super-resolved method to the projected DNS solution for initial condition $u(x,0)=2.5e^{-20(x-\pi)^2}$.}
\label{AD_online2}
\end{figure}

\begin{figure}[h]
\centering
\includegraphics[width=1.0\textwidth]{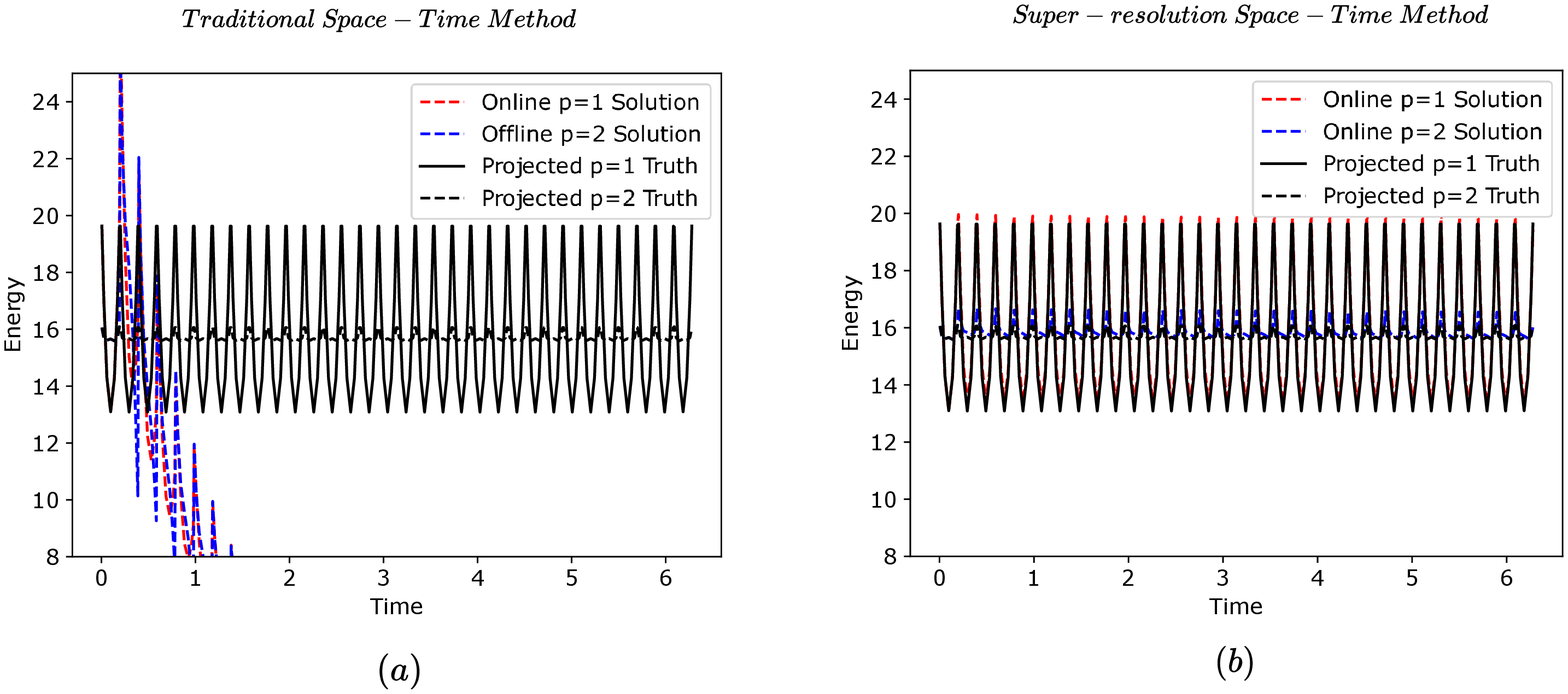}
\caption{Evolution of energy $E(t) = \int u(x,t)^2 d\Omega_x$ as a function of time for traditional space-time method vs. super-resolution method for initial condition $u(x,0)=sin(4x)+2sin(8x)$.}
\label{AD_online_energy}
\end{figure}

\begin{figure}[h]
\centering
\includegraphics[width=1.0\textwidth]{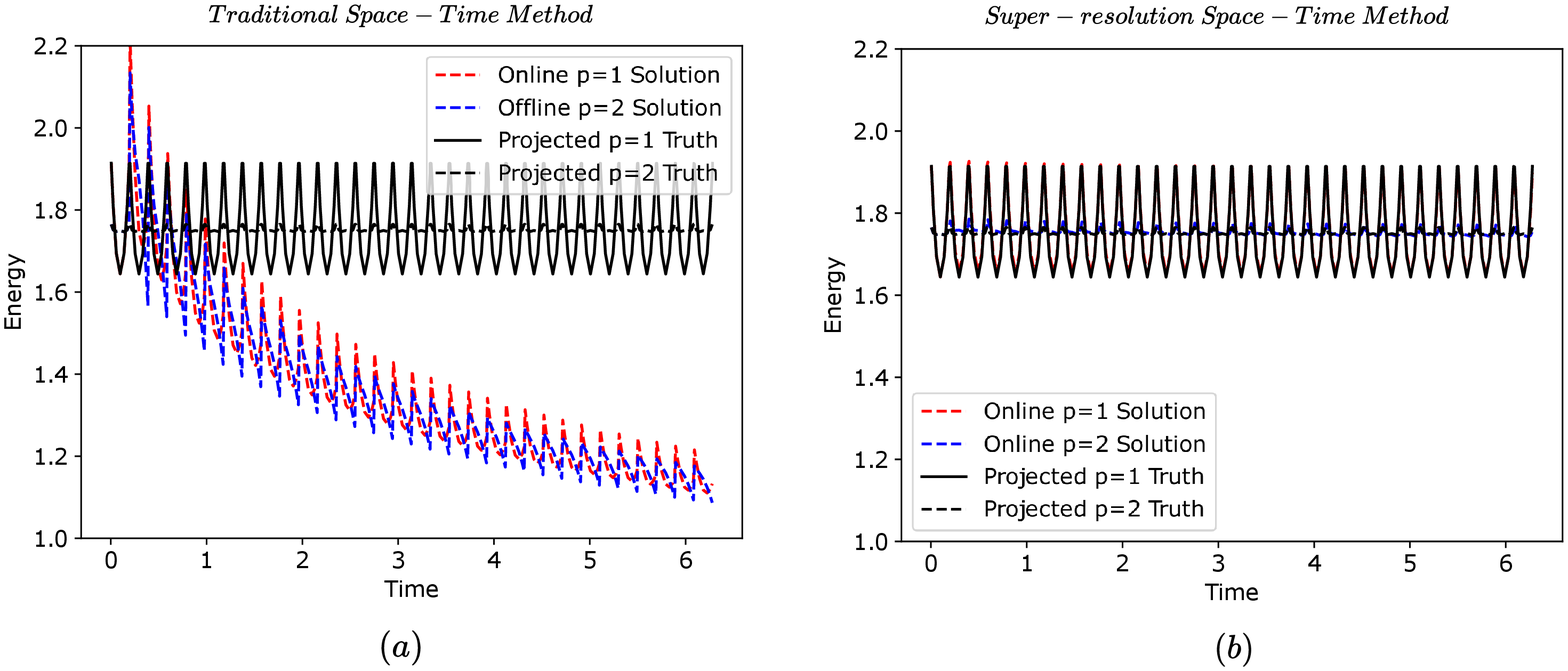}
\caption{Evolution of energy $E(t) = \int u(x,t)^2 d\Omega_x$ as a function of time for traditional space-time method vs. super-resolution method for initial condition $u(x,0)=2.5e^{-20(x-\pi)^2}$.}
\label{AD_online_energy2}
\end{figure}

\begin{table}[]
\begin{adjustbox}{width=\textwidth}
\begin{tabular}{lllllllllll}
$N_x$ & $N_t$ & $\Delta x$ & $\Delta t$ & $CFL$ & $\lVert u_2-u_1\lVert_2$ & $\lVert u_{2,NN}-u_1\lVert_2$ & $\lVert u_{2,NN}-u_{2}\lVert_2$ & $\lVert u_2-u_1\lVert_2 \over \lVert u_2\lVert_2$ & $\lVert u_{2,NN}-u_1\lVert_2 \over \lVert u_2\lVert_2$ & $\lVert u_{2,NN}-u_{2}\lVert_2 \over \lVert u_2\lVert_2$ \\ 
\hline 
16 & 32 & 0.3927 & 0.19635 & 0.5 & 0.41684 & 0.41684 & 0.0018975 & 0.066344 & 0.066345 & 0.00030201 \\ 
16 & 16 & 0.3927 & 0.3927 & 1 & 0.56984 & 0.57053 & 0.0044046 & 0.090699 & 0.090809 & 0.00070106 \\ 
16 & 8 & 0.3927 & 0.7854 & 2 & 1.4709 & 1.4681 & 0.0066213 & 0.23456 & 0.23411 & 0.0010559 \\ 
16 & 4 & 0.3927 & 1.5708 & 4 & 3.3222 & 3.3223 & 0.016548 & 0.56392 & 0.56394 & 0.0028089 \\ 
16 & 2 & 0.3927 & 3.1416 & 8 & 3.115 & 2.9975 & 1.7061 & 0.75135 & 0.723 & 0.41152 \\ 
32 & 64 & 0.19635 & 0.098175 & 0.5 & 0.1075 & 0.10755 & 0.00075342 & 0.01711 & 0.017117 & 0.00011991 \\ 
32 & 32 & 0.19635 & 0.19635 & 1 & 0.14736 & 0.14738 & 0.0020468 & 0.023453 & 0.023457 & 0.00032576 \\ 
32 & 16 & 0.19635 & 0.3927 & 2 & 0.41684 & 0.4157 & 0.005024 & 0.066344 & 0.066163 & 0.00079962 \\ 
32 & 8 & 0.19635 & 0.7854 & 4 & 1.4235 & 1.4245 & 0.009266 & 0.22699 & 0.22716 & 0.0014776 \\ 
32 & 4 & 0.19635 & 1.5708 & 8 & 3.3146 & 2.723 & 1.3326 & 0.5626 & 0.46219 & 0.2262 \\ 
\hline 
\end{tabular}
\end{adjustbox}
\centering
\caption{Reconstruction error when super-resolved from $p=1$ to $p=2$ for the linear advection problem with an unseen initial condition at different CFL numbers.}
\label{table12}
\end{table}

\begin{table}[]
\begin{adjustbox}{width=\textwidth}
\begin{tabular}{lllllllllll}
$N_x$ & $N_t$ & $\Delta x$ & $\Delta t$ & $CFL$ & $\lVert u_3-u_1\lVert_2$ & $\lVert u_{3,NN}-u_1\lVert_2$ & $\lVert u_{3,NN}-u_{3}\lVert_2$ & $\lVert u_3-u_1\lVert_2 \over \lVert u_3\lVert_2$ & $\lVert u_{3,NN}-u_1\lVert_2 \over \lVert u_3\lVert_2$ & $\lVert u_{3,NN}-u_{3}\lVert_2 \over \lVert u_3\lVert_2$ \\ 
\hline 
16 & 64 & 0.3927 & 0.098175 & 0.25 & 0.40799 & 0.40956 & 0.029322 & 0.064934 & 0.065183 & 0.0046667 \\ 
16 & 32 & 0.3927 & 0.19635 & 0.5 & 0.42022 & 0.42039 & 0.004307 & 0.066879 & 0.066907 & 0.00068548 \\ 
16 & 16 & 0.3927 & 0.3927 & 1 & 0.57471 & 0.57558 & 0.0043912 & 0.091468 & 0.091606 & 0.00069888 \\ 
16 & 8 & 0.3927 & 0.7854 & 2 & 1.5204 & 1.5228 & 0.0087391 & 0.242 & 0.24238 & 0.001391 \\ 
16 & 4 & 0.3927 & 1.5708 & 4 & 3.8684 & 3.8747 & 0.018305 & 0.62236 & 0.62338 & 0.0029449 \\ 
16 & 2 & 0.3927 & 3.1416 & 8 & 3.8513 & 3.6725 & 3.079 & 0.81524 & 0.77739 & 0.65176 \\ 
32 & 128 & 0.19635 & 0.049087 & 0.25 & 0.108 & 0.10626 & 0.0096799 & 0.017188 & 0.016912 & 0.0015406 \\ 
32 & 64 & 0.19635 & 0.098175 & 0.5 & 0.10772 & 0.10788 & 0.0011823 & 0.017145 & 0.017169 & 0.00018816 \\ 
32 & 32 & 0.19635 & 0.19635 & 1 & 0.14764 & 0.14791 & 0.0015759 & 0.023498 & 0.02354 & 0.00025081 \\ 
32 & 16 & 0.19635 & 0.3927 & 2 & 0.42022 & 0.42123 & 0.0037693 & 0.066879 & 0.06704 & 0.0005999 \\ 
32 & 8 & 0.19635 & 0.7854 & 4 & 1.4737 & 1.4799 & 0.011139 & 0.23456 & 0.23554 & 0.0017729 \\ 
32 & 4 & 0.19635 & 1.5708 & 8 & 3.8616 & 3.0788 & 1.266 & 0.62126 & 0.49532 & 0.20368 \\ 
\hline 
\end{tabular}
\end{adjustbox}
\centering
\caption{Reconstruction error when super-resolved from $p=1$ to $p=3$ for the linear advection problem with an unseen initial condition at different CFL numbers.}
\label{table13}
\end{table}

\subsection{Super-resolution}
To train the super-resolution model, we will generate the true solution on a fine grid. The initial condition for this case is $sin(x)+sin(2x)+sin(4x)$, and the size of the grid is taken to be $N_x \times N_t:512 \times 512$.  This high-resolution mesh is to ensure that the true solution remains highly resolved on this grid. For simplicity, periodic boundary conditions are also applied in the $x$ direction. The true solution is then evaluated on all the grid points to obtain the DNS solution. The next step is to obtain the coarse $p=1$ and fine $p=2$ $L_2$-projected solution for different meshes having spatial and temporal element sizes $\Delta x$ and $\Delta t$ respectively as shown in figure \ref{l2proj_ad}. A non-dimensional parameter naturally arising in this case is the $CFL = {a \Delta t \over \Delta x}$ number which is similar to Peclet number in the 1-D convection-diffusion problem. This solution is then projected on grids of various sizes corresponding to CFL numbers of 0.25,0.5,1.0,2 and 4. These CFL numbers correspond to three sets of grid-sizes: (i.) $N_x \times N_t:$ 32$\times$128, 32$\times$64, 32$\times$32, 32$\times$16 and 32$\times$8; (ii.) $N_x \times N_t:$ 16$\times$64, 16$\times$32, 16$\times$16, 16$\times$8 and 16$\times$4; and (iii.) $N_x \times N_t:$ 24$\times$96, 24$\times$48, 24$\times$24, 24$\times$12 and 24$\times$6. For each element in these grids, the basis coefficients corresponding to the coarse-space is extracted along with its neighbors, excluding those which are part of the future time step, i.e., only space-time elements present in the south, south-east, south-west, east, and west of the central coarse element are extracted. The corresponding fine-space basis coefficients are also extracted for the central element. As a first step towards normalization, a mean value $u_m$ is first computed inside the element as follows:
\begin{equation}
   u_m = {\iint_{\Omega_e} u_1(x,t) \,dx\,dt \over \iint_{\Omega_e} \,dx\,dt}.
\end{equation}
Similarly, an R.M.S value is also computed inside the element:
\begin{equation}
   u_{rms} = \sqrt{\iint_{\Omega_e} (u_1(x,t)-u_m)^2 \,dx\,dt \over \iint_{\Omega_e} \,dx\,dt}.
\end{equation}
A model is then sought in the following form
\begin{align}
    \left[C'_{1,p_s},C'_{2,p_s},..,C'_{p_s+1,p_s}\right] = \mathbf{f}\biggr(log(CFL),\left[\tilde{C}_{1,p_c},\tilde{C}_{2,p_c},..,\tilde{C}_{p_c+1,p_c}\right],..\\
    ,\left[\tilde{C}_{1,p_c},\tilde{C}_{2,p_c},..,\tilde{C}_{p_c+1,p_c}\right]_m,..\biggr), \nonumber
     \label{fform}
\end{align}
where $C'_{i,j}$ and $\tilde{C}_{i,j}$ are defined in terms of $u_m$ and $u_{rms}$ similar to equation \eqref{basic_frm} . For learning this function, the VSRNN architecture is adopted with size 24$\times$12 for part A, 1$\times$6$\times$12 for part B, and 12$\times$9 for the post-multiplication part, respectively. The network is first trained on data obtained by projecting a highly resolved DNS solution, as shown in figure \ref{l2proj_ad}.  

In the next step, the super-resolution model is evaluated on an unseen coarse solution obtained by projecting the DNS solution for a different set of initial conditions, as shown in figure \ref{ad_offline_12}. It can be observed that unless an extremely coarse model was used, the model could efficiently super-resolve unseen coarse solution to its fine-solution with minimal reconstruction error. This reconstruction error for the cases is reported in table \ref{table12}. Except for the case when the CFL number was as high as 8, the error in reconstruction $\lVert u_{2,NN}-u_{2}\lVert_2$ is orders of magnitude smaller in comparison to the magnitude of the sub-scale $\lVert u_{2}-u_{1}\lVert_2$. Hence, the super-resolution model is very efficient in reconstructing the fine-solution as long as the CFL is not large. A separate model for super-resolving $p=1$ solution to $p=3$ solution is also trained by repeating the steps above. As reported in table \ref{table13}, a similar trend is again observed for this case where efficient reconstruction was obtained at lower CFL values. Irrespective of the large reconstruction error at CFL values of 8 and above, the magnitude of $\lVert u_3-u_1\lVert_2$  is still very close to $\lVert u_{3,NN}-u_1\lVert_2$ for all CFL numbers. This indicates that the super-resolution model can also act as an efficient error indicator that can be used for mesh adaption. 

\subsection{Sub-grid Modelling}
In the previous section, we showed that the neural network could efficiently predict the sub-scales as long as the grid is not highly under-resolved. In this section, we will use the trained model from the previous section to improve the existing space-time based numerical method. To this end, we start with the linear advection equation in the domain $\Omega \subset \mathbb{R}$ with the boundary $\Gamma = \partial \Omega$ as follows
\begin{equation}
    \frac{\partial u}{\partial t} + a  \frac{\partial u}{\partial x} = 0,
\end{equation}
with a periodic boundary condition at the boundary $\Gamma$ and time $t \in ( 0,T \rbrack$. The weak form of the above equation is obtained by multiplying it with a test function $w$ and integrating it over the space-time element as follows
\begin{equation}
    \int_{\Omega_e}\left(\frac{\partial u}{\partial t} + a  \frac{\partial u}{\partial x}\right){w}{d\Omega} = 0,
\end{equation}
such that $u \in \mathcal{V}$ for all $w \in \mathcal{V}$. Let us also define $\mathcal{T}_{h}$ as a tessellation of the domain $\Omega$ into a set of non-overlapping elements, $K$, each having a sub-domain $\Omega_e$ and boundary $\Gamma_e$. The vector space of functions $\mathcal{V}\equiv\mathcal{H}^1(\mathcal{T}_{h})$ is a  Sobolev space where the functions and their derivatives are square-integrable {\em inside each element}. Simplifying the previous equation, we obtain the following:
\begin{equation}
    \int_{\Omega_e}\left(\frac{\partial uw}{\partial t} + \frac{\partial auw}{\partial x}\right){d\Omega} -  \int_{\Omega_e}\left(u\frac{\partial w}{\partial t} + au\frac{\partial w}{\partial x}\right){d\Omega}= 0.
\end{equation}
Application of the divergence theorem leads to the following
\begin{equation}
     \int_{\Gamma_e}(auw \mathbf{\hat{i}} + uw \mathbf{\hat{j}}).(n_x \mathbf{\hat{i}} + n_t \mathbf{\hat{j}})d{\Gamma} -  \int_{\Omega_e}\left(u\frac{\partial w}{\partial t} + au\frac{\partial w}{\partial x}\right){d\Omega}= 0,
\end{equation}
where $n_x$ and $n_t$ denote the components of the outward normal on the surface of the element along the space and time axis, respectively. The first term in the DG method is replaced with a numerical flux as follows:
\begin{equation}
     \int_{\Gamma_e}(F_x^*(au,au^-) \mathbf{\hat{i}} + F_t^*(u,u^-) \mathbf{\hat{j}}).(n_x \mathbf{\hat{i}} + n_t \mathbf{\hat{j}})wd{\Gamma} -  \int_{\Omega_e}\left(u\frac{\partial w}{\partial t} + au\frac{\partial w}{\partial x}\right){d\Omega}= 0.
     \label{dg_strong}
\end{equation}
The traditional space-time DG method can be obtained by applying the Galerkin approximation to the previous equation as follows:
\begin{equation}
     \int_{\Gamma_e}(F_x^* \mathbf{\hat{i}} + F_t^* \mathbf{\hat{j}}).(n_x \mathbf{\hat{i}} + n_t \mathbf{\hat{j}}){w_h}d{\Gamma} -  \int_{\Omega_e}\left({u_h}\frac{\partial {w_h}}{\partial t} + a{u_h}\frac{\partial u_h}{\partial x}\right){d\Omega}= 0,
\end{equation}
where $F_x^* = a \tilde{u}^{-}$ when $a n_x<0$ and $a \tilde{u}$ when $an_x>0$. {\color{black}The temporal flux on the bottom face is based on previous space-time slab i.e. $F_t^* = \tilde{u}^-$}. The effect of the numerical fluxes is similar to that of a closure, which is dissipative in action due to the jump term, ensuring the stability of the method. The numerical fluxes were originally developed for solving the exact 1-D problem at the interface, and application to the DG method is generally made by applying it along the normal direction of the interface. However, this might not be the most optimal choice for the flux. To this end, we revisit the strong form of the DG i.e., equation \eqref{dg_strong} through the VMS approach. The coarse-scale equation corresponding to equation \eqref{dg_strong} is given by:
\begin{equation}
     \int_{\Gamma_e}\left(F_x^*(a(u_h + u'),a(u_h^{-} + u'^{-})) \mathbf{\hat{i}} + F_t^*(u_h + u',u_h^{-} + u'^{-}) \mathbf{\hat{j}}\right).(n_x \mathbf{\hat{i}} + n_t \mathbf{\hat{j}})w_h d{\Gamma} -  \int_{\Omega_e}\left((u_h + u')\frac{\partial w_h}{\partial t} + a(u_h + u')\frac{\partial w_h}{\partial x}\right){d\Omega}= 0.
\end{equation}
If  $\mathcal{V}_h \bot \mathcal{V}'$, then the effect of the sub-scale on the interior flux is negligible i.e.
\begin{equation}
      \int_{\Gamma_e}\left(F_x^*(a(u_h + u'),a(u_h^{-} + u'^{-})) \mathbf{\hat{i}} + F_t^*(u_h + u',u_h^{-} + u'^{-}) \mathbf{\hat{j}}\right).(n_x \mathbf{\hat{i}} + n_t \mathbf{\hat{j}})w_h d{\Gamma} -  \int_{\Omega_e}\left(u_h\frac{\partial w_h}{\partial t} + a u_h\frac{\partial w_h}{\partial x}\right){d\Omega}= 0,
\end{equation}
and the effect of un-resolved sub-scales is only through the flux. The true solution $u_h + u'$ is infinite-dimensional. However, only a few of its moments are required in the form of inner-products with low-order basis functions on element faces.  In this limit, we assume that the following approximation can be made: $u_s \approx u_h + u'$, where $u_s$ denotes the super-resolved solution of $u_h$ i.e. 
\begin{equation}
	\begin{aligned}
		\int_{\Gamma_e}\left(F_x^*(a(u_h + u'),a(u_h^{-} + u'^{-})) \mathbf{\hat{i}} + F_t^*(u_h + u',u_h^{-} + u'^{-}) \mathbf{\hat{j}}\right).(n_x \mathbf{\hat{i}} + n_t \mathbf{\hat{j}})w_h d{\Gamma}  \\
		\approx \int_{\Gamma_e}\left(F_x^*(au_s,au^-_s) \mathbf{\hat{i}} + F_t^*(u_s,u_s^-) \mathbf{\hat{j}}\right).(n_x \mathbf{\hat{i}} + n_t \mathbf{\hat{j}})w_h d{\Gamma},
	\end{aligned}
\end{equation}
where $F_x^*$ and $F_t^*$ are traditional up-wind numerical fluxes but computed using the super-resolved state $u_s$ instead of $u_h$. In this paper, we choose the spaces of $u_h$ and $u_s$ to be $p=1$ and $p=2$ respectively. 

To obtain the super-resolved state, we will re-use the super-resolution network trained in the previous {\color{black}subsection}. Two different initial conditions are chosen:{\color{black} (i.) Case A with an initial condition $sin(4x)+2sin(8x)$ (ii.) Case B with an initial condition $2.5e^{-20(x-\pi)^2}$}. These initial conditions are different from those used for training and testing. A comparison of different initial conditions used in training, reconstruction, and online evaluation is summarised in figure \ref{init_cond}. The space-time slab is then discretized into 32 elements in the spatial direction, and the CFL value is taken to be 1.0. Figure \ref{AD_online} and \ref{AD_online2} shows results obtained for the super-resolution model and the traditional model compared to the optimal solution obtained by $L_2$-projection of the DNS solution for the two different initial conditions, respectively. The super-resolution model is far more accurate than the traditional method, where the sub-scales were recovered with a high level of accuracy for both cases. In the case of the traditional method, extrema can be seen decreasing with time considerably in comparison to the super-resolution space-time method both in figures \ref{AD_online} and \ref{AD_online2}. This shows a higher dissipation characteristic of the traditional numerical method over the super-resolution method. This can also be quantitatively seen in both the figures \ref{AD_online_energy}(a) and \ref{AD_online_energy2}(a), where the red line denoting time evolution of energy i.e. {\color{black} $E(t) = \int_{\Omega_x} u_h(x,t)^2 d\Omega_x$ decreases in time for the traditional approach in comparison to the optimal $p=1$ solution which is oscillatory but energy conservative. Hence, the optimal solution oscillates about a fixed value because the true solution itself conserves energy. The reason $E(t)$ oscillates for the optimal solution is because it is computed by integrating $u_h(x,t)^2$ over only space $x$ and can still vary in time across the space-time slab.}

In the next step, we obtain the space-time solution by first solving the problem using the traditional approach and applying our super-resolution model on this coarse solution. In this case, the super-resolution network has no contribution to the numerical simulation stage. Rather, it is used offline when the solution is made available. As shown in figure \ref{AD_online_energy}(a) and \ref{AD_online_energy2}(a), application of the super-resolution in an offline stage does not improve the results. On the other hand, when the super-resolved states were used to compute the flux in the numerical method online, a high level of $L_2$-optimality was also obtained in the coarse solution as shown in figures \ref{AD_online_energy}(b) and \ref{AD_online_energy2}(b). Consequently, the corresponding super-resolved solution was also accurate and close to the $p=2$ optimal solution. 

\begin{figure}[h]
\centering
\includegraphics[width=1.0\textwidth]{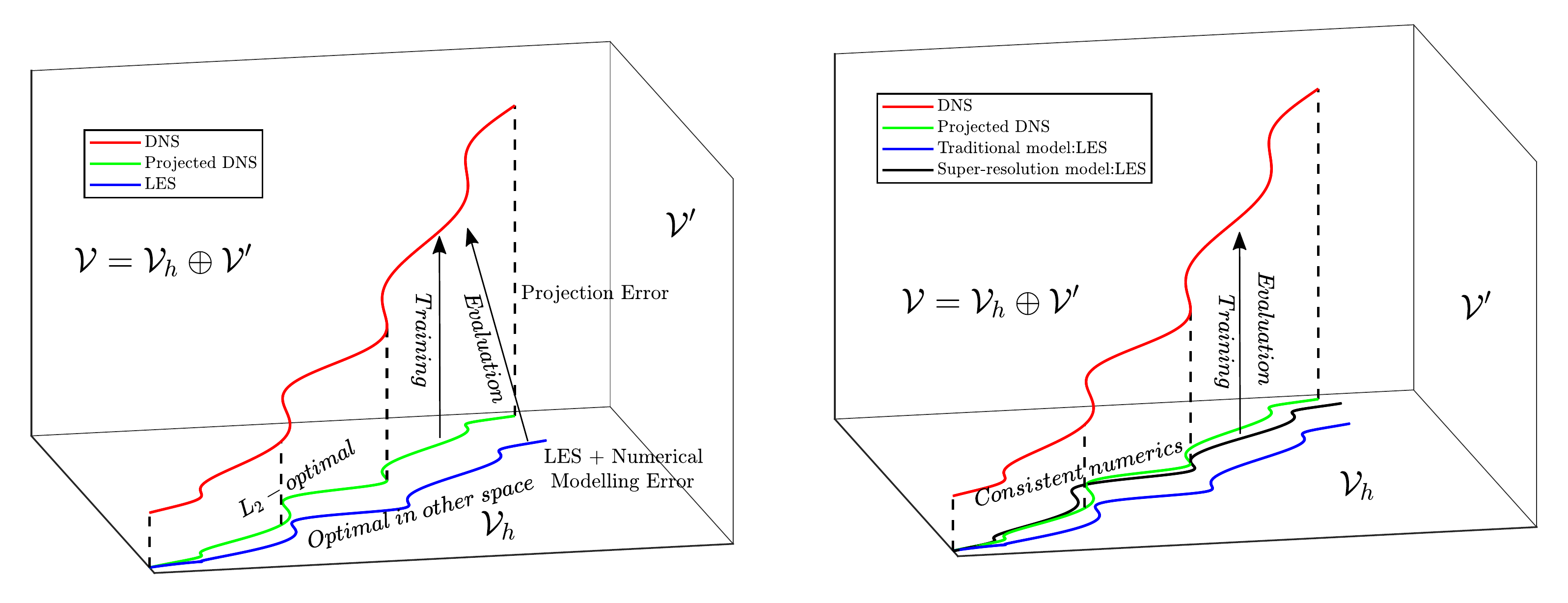}
\caption{Sources of errors in offline and online super-resolution.}
\label{offlineonlline}
\end{figure}

As observed previously in this section, the super-resolution did not improve the result when it was used on the data obtained using the traditional DG method. However, when used in an online setting, it improved the performance of the numerical method. This can be explained by figure \ref{offlineonlline}, which shows the evolution of the DNS solution  (red line) in an infinite-dimensional space. When one trains the super-resolution model, the mapping from a point on the green line (optimal LES) to its corresponding point on the red line (DNS) is learnt. However, when running an online numerical simulation using a traditional approach, the trajectory (blue line) taken by the solution (LES) is entirely different. The model encounters input parameters that it has not encountered during training and outputs an incorrect super-resolved solution in the evaluation stage. The blue line  represents another optimal representation of DNS solution on the coarse solution space and not the $L_2$-optimal solution for which the method has been trained. However, when the super-resolved state is used to compute the fluxes, it forces the coarse resolution towards $L_2$-optimality because the closure has been formulated using the VMS method, where the coarse and fine spaces are formally defined.  As shown in the second part of figure \ref{offlineonlline}, consistent numerical methods are required for the super-resolution models to work correctly. The VMS method is an ideal candidate to help us in achieving this consistency.  

\section{p-Super-resolution of turbulent channel flow.}\label{sec:turbchan}
The assessment of super-resolution models for turbulent flows poses a stern challenge. {\color{black}This is because, given limited measurements from a severely under-resolved coarse-grained solution such as LES, there are infinitely many possible solutions for the fine-scales.} This problem is especially true for filtering using the sharp spectral filter, where the fine-scale solution is lost after filtering, and it is impossible to recover the original field from filtered data. The exact fine-scales are both functions of the coarse space and their time-history \cite{STINIS,MZ1,MZ2,MZVMS, pradhan2019variational}. As shown by Langford and Moser \cite{optimal} in their work on optimal LES, to compute the correct single-time multi-point statistical quantity of the large-scale field exact fine-scales might not be required. The Smagorinsky and the VMS models \cite{VMS,OSS,OSS2,VMS3,VMSE,NLVMS2,pradhan2019variational}, which perform well online, are well-known to perform poorly in an a priori setting. Similarly, the N-N generated super-resolved field is not point-wise exact. Rather it is an optimal representation of the fine-space generating correct single-time multi-point statistics. 

Our model being compact, the $L_2$ error is computed only locally in a single element. The training data consists of data from several elements, part of a 2-D DNS slice having homogeneity in stream-wise and span-wise directions. During the optimization, these errors from each element are averaged. As a result, the model output, in some sense, is an optimal representation of the fine-scales for all possible realizations. To this end, we will be using one-dimensional energy spectra that have been averaged over homogeneous directions as a measure for model accuracy in-place of the $L_2$ norm for the full field. The contours of the reconstructed fine-space solution will only be presented for qualitative purposes.

The first step is to generate data for training the model. As described in section 4, a single 2-D DNS snapshot is extracted at a wall-normal height of $y^+ \approx 850$ and is $L_2$-projected on discontinuous polynomial spaces spanned by order $p=1,3$ tensor-product Lagrange basis functions on meshes of different sizes. In this case, we project the DNS solution on meshes with elements $N_x \times N_y$: 8$\times$4, 16$\times$8, 32$\times$16, 64$\times$32 in the $x$ and $y$ directions respectively. Once the $p=1,3$ projected solutions are obtained, for each element present in these meshes, $p=1$ coarse-scale basis coefficients are extracted for both the element and its immediate neighbors along with the $p=3$ fine-scale basis coefficients. The next step is to evaluate the normalising parameter for each element i.e. $u_{rms} = \sqrt{\int_{\Omega_e} (u_1 - u_m)^2  d\Omega_e \over \int_{\Omega_e} d\Omega_e}$, where the mean velocity $u_m$ inside an element is given by $u_m = {\int_{\Omega_e} u_1  d\Omega_e \over \int_{\Omega_e} d\Omega_e }$. Finally, a functional form similar to equation \eqref{fform} is assumed except the parameter $\alpha$ is replaced with the logarithm of cell Reynolds number i.e. $log(Re_{\Delta})$. The physics-informed feature $log(Re_{\Delta})$ ensures that different orders of magnitude of $Re_{\Delta}$ in training data is accounted for. Finally, the normalized input and output basis coefficient data and the logarithmic cell Reynolds number $log(Re_{\Delta})$ for each element are assembled into a single table as a training data-set. 

\begin{figure}[h]
\centering
\includegraphics[width=1.0\textwidth]{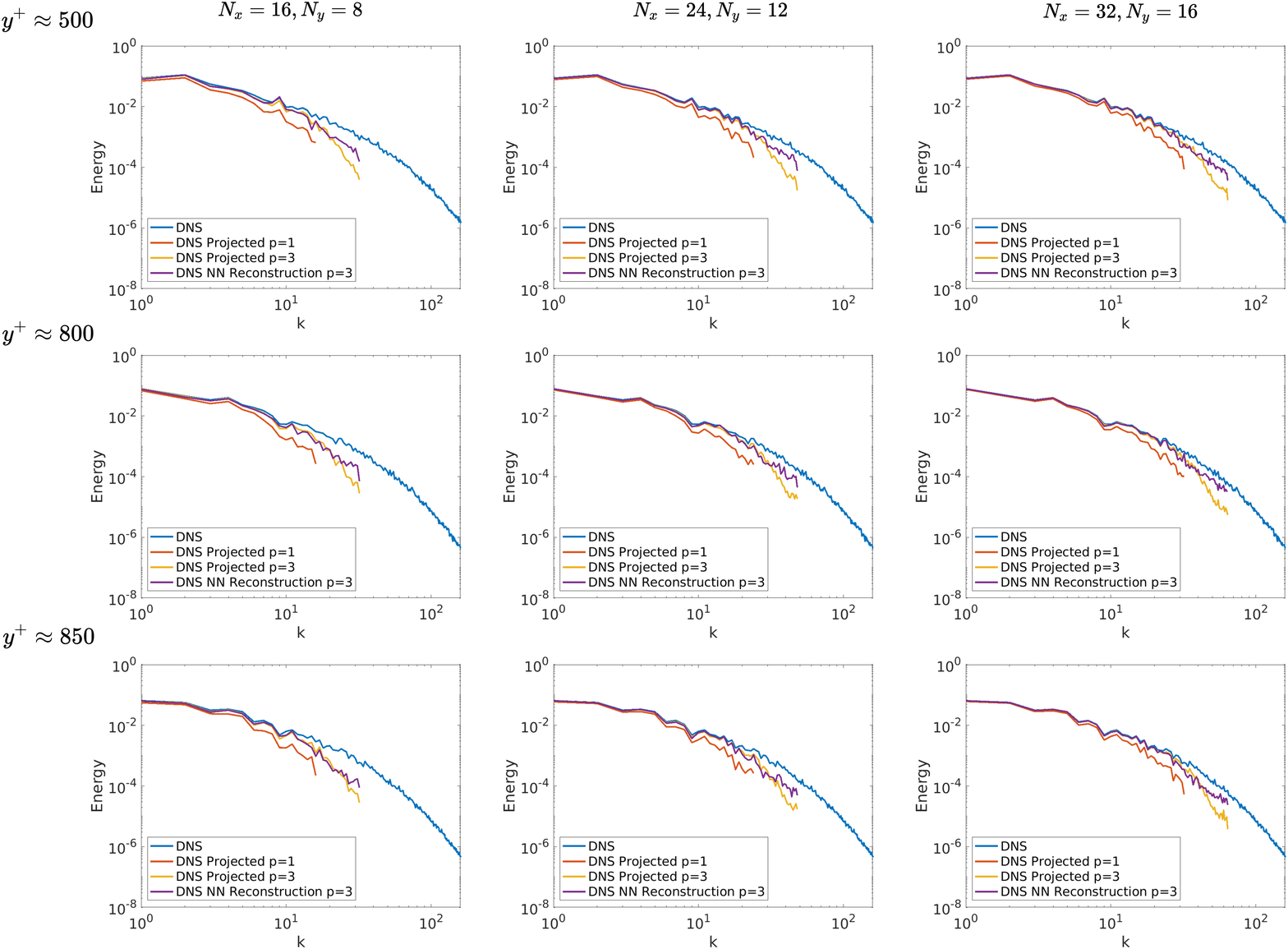}
\caption{Stream-wise energy spectra obtained for the $L_2$-projected stream-wise velocity solution on $p=1$, $L_2$-projected stream-wise velocity solution on $p=3$, N-N super-resolved $p=3$ solution and DNS at different wall normal height $y+$ and mesh resolutions.}
\label{stream_turb}
\end{figure}

\begin{figure}[h]
\centering
\includegraphics[width=1.0\textwidth]{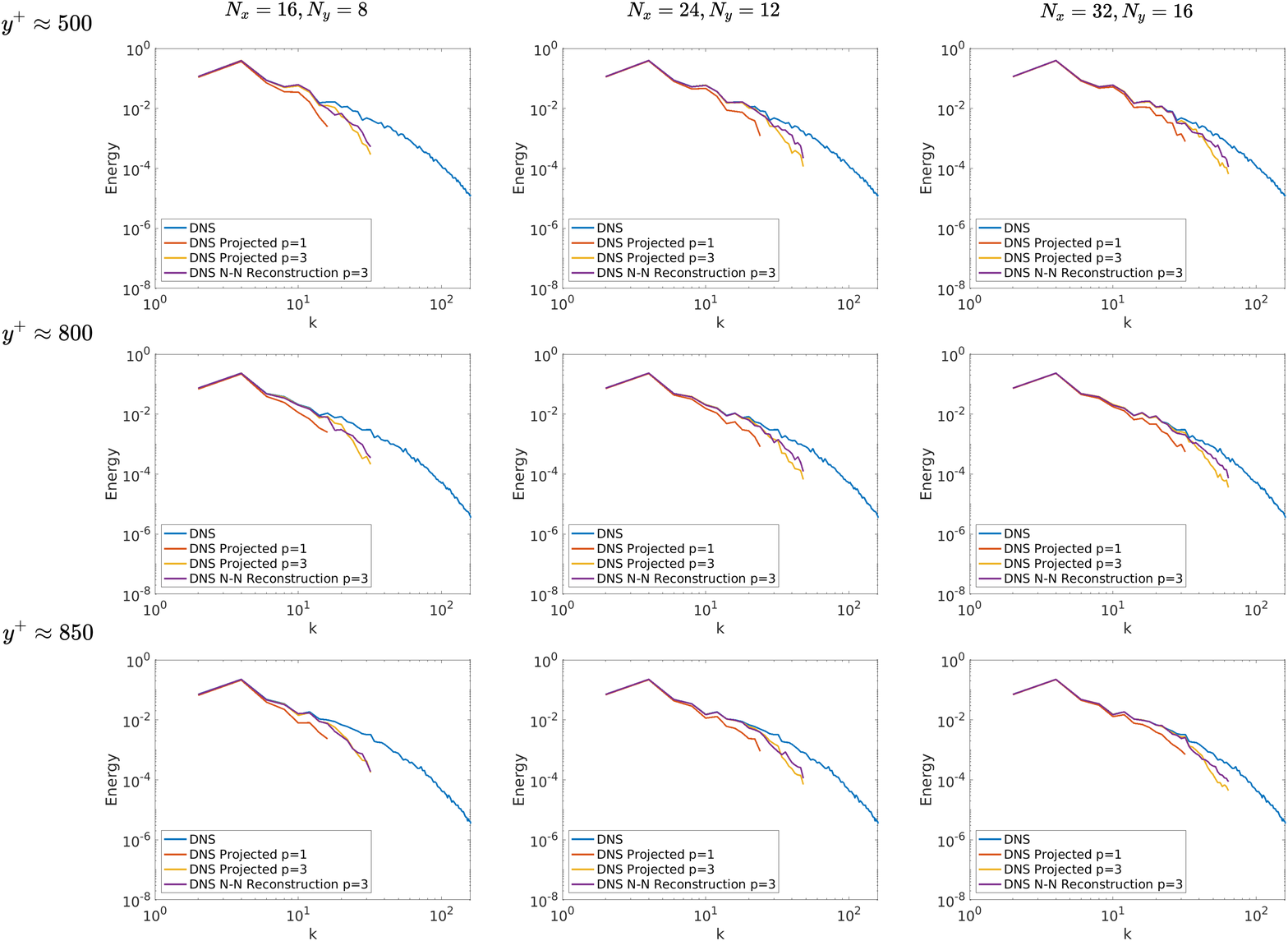}
\caption{Span-wise energy spectra obtained for the $L_2$-projected stream-wise velocity solution on $p=1$, $L_2$-projected stream-wise velocity solution on $p=3$, N-N super-resolved $p=3$ solution and DNS at different wall normal height $y^+$ and mesh resolutions.}
\label{span_turb}
\end{figure}

A VSRNN architecture for the N-N model is then assumed with sizes: 37$\times$32$\times$32$\times$32 for the part A, 16$\times$32$\times$32 for the part B, and a 32$\times$64$\times$64$\times$16 sized post-multiplication part, respectively. Finally, the model performance is evaluated in figures \ref{span_turb} and \ref{stream_turb} by comparing the stream-wise and span-wise energy spectra obtained for the super-resolved $p=3$ solution and the $p=3$ $L_2$-optimal solution at different wall-normal heights of $y^+ \approx 500,800, 850$. To compute the energy spectra, the solution in first obtained on a uniform mesh with size $(p+1)N_x \times (p+1)N_y$ where the factor $p+1$ accounts for the effective grid-size at higher orders. This also prevents the over-sampling of the data. 

\begin{figure}[h]
\centering
\includegraphics[width=0.80\textwidth]{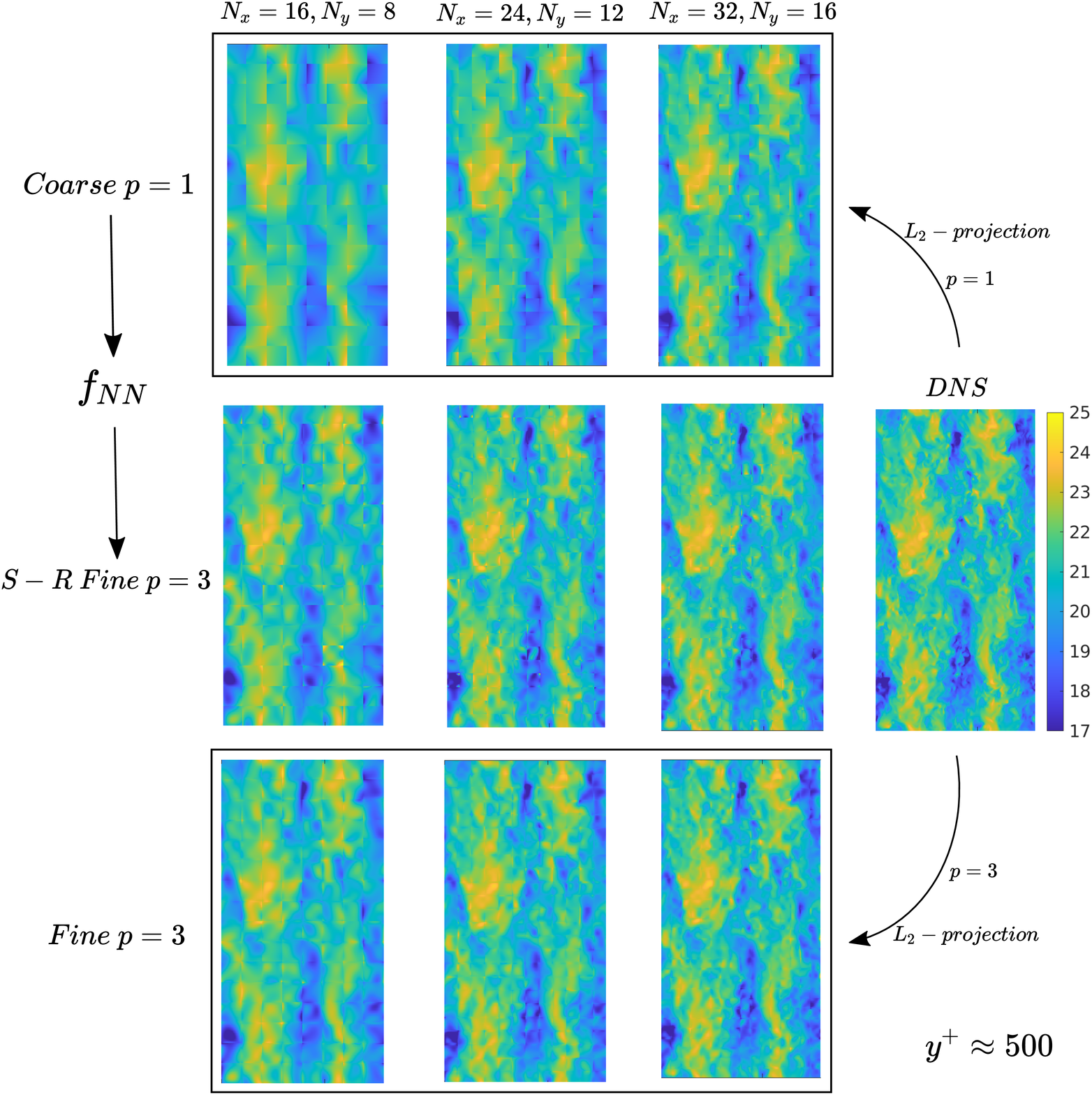}
\caption{Stream-wise Velocity contours of the coarse solution, super-resolved fine solution and the corresponding optimal fine solution.}
\label{qual_super}
\end{figure}

As can be observed in figures \ref{span_turb} and \ref{stream_turb}, the network can successfully recreate the correct energy distribution across different wave-numbers both in the stream-wise and the span-wise directions. This is true for both the cases: the plane at $y^+\approx 850$, which was used for training, and the unseen planes at $y^+\approx 500$ and $y^+\approx 800$. However, the energy at the high wave-number modes for all these cases was slightly higher than the $L_2$-projected $p=3$ optimal solution suggesting that a small amount of de-aliasing would be helpful. A qualitative plot of the coarse $p=1$ solution, the super-resolved $p=3$ solution and the $L_2$-optimally projected $p=3$ solution for different grid sizes at a wall-normal distance of $y^+\approx500$ is shown in figure \ref{qual_super}. It can be observed that the super-resolved solution, similar to the optimal $p=3$ projected solution, has finer structures when compared to the coarse $p=1$ solution at different mesh resolutions.

The generalizability of the model trained using a single snapshot of DNS data stems from the fact that when the DNS image is projected on several finite element meshes with different element sizes, the average value of the cell Reynolds number changes. As a result, the training data contains an extensive range of cell Reynolds numbers. When the trained model is evaluated at different wall-normal distances while retaining the same the grid size, the cell Reynolds number changes due to changes in $u_{rms}$. However, this new cell Reynolds number can also be obtained at a previous wall-normal height by changing the grid-size alone. This can also be observed in equation \eqref{para_dep} for the normalized sub-scales. The normalized sub-scales only depend on the cell Peclet number $\alpha$ and the non-dimensionalized inputs rather than the grid size or the diffusion coefficient separately.

\subsection{Sub-grid Modelling}
The previous {\color{black}subsection} showed that the super-resolution model accurately reconstructs the high-order optimal solution from the low-order optimal solution. 
We consider LES of the compressible Navier--Stokes equation, but for simplicity of presentation, we only detail the development for the inviscid terms. The domain $\Omega \subset \mathbb{R}^d$ is used with the boundary $\Gamma = \partial \Omega$, where $d\geq1$ is the dimension of the problem as follows,
\begin{equation}
    {\partial{\mathbf{u}} \over \partial t} + \nabla\cdot{\mathbf{F(u)}}=  0,
\end{equation}
with appropriate boundary conditions on $\Gamma$ and time $t \in ( 0,T \rbrack$. The state vector $\mathbf{u}$ is given by:
\begin{equation}
    {\mathbf{u}} =  [
    \rho,\rho u,\rho v,\rho w,\rho E]^{T},
\end{equation}
and the matrix $\mathbf{F(u)}$ corresponding to the flux is given by:
\begin{equation}
\mathbf{F(u)} = 
\begin{bmatrix}
    \rho u & \rho v & \rho w  \\
    \rho u^2+p & \rho uv & \rho uw\\
    \rho vu & \rho v^2+p & \rho vw\\
    \rho wu & \rho wv & \rho w^2+p\\
    \rho uH & \rho vH & \rho wH  \\
\end{bmatrix}
.
\end{equation}
Let us also define $\mathcal{T}_{h}$ as a tessellation of the domain $\Omega$ into a set of non-overlapping elements, $K$, each having a sub-domain $\Omega_k$ and boundary $\partial \Omega_k$. The DG weak form is then obtained by multiplying with weighting functions $\mathbf{w}$ and performing integration by parts on an element:

\begin{equation}
    \int_{\Omega_k}\left({\partial{\mathbf{u}} \over \partial t} + \nabla\cdot \mathbf{F(u)}\right)\cdot\mathbf{w} d\Omega=  0.
\end{equation}
After application of integration by parts we obtain:
\begin{equation}
    \int_{\Omega_k}{\partial{\mathbf{u}} \over \partial t}\cdot\mathbf{w}d\Omega-  \int_{\Omega_k}{\nabla \mathbf{w}: \mathbf{F(u)}}d\Omega + \int_{\partial \Omega_k}{ \mathbf{w}\cdot \mathbf{F^{*}(u,u^{-})}}d\Gamma =  0.
\end{equation}
Using the Galerkin approximation we have, 
\begin{equation}
    \int_{\Omega_k}{\partial{\mathbf{u_h}} \over \partial t}\cdot\mathbf{w_h}d\Omega-  \int_{\Omega_k}{\nabla \mathbf{w_h}: \mathbf{F(u_h)}}d\Omega + \int_{\partial \Omega_k}{ \mathbf{w_h}\cdot \mathbf{F^{*}(u_h,u_h^{-})}}d\Gamma =  0,
\label{ILES}
\end{equation}
where components of $\mathbf{u_h}$, i.e., ${u_{h,i}} \in \tilde{\mathcal{V}_3}$ and $\mathcal{V}_3$ is the space of $p=3$ tensor-product polynomial basis functions as follows:
\begin{equation}
\tilde{\mathcal{V}_3} \triangleq \left\{u\in{L_2}(\Omega):u|_{\mathcal{T}}\in P^3(T),T\in\mathcal{T}_{h} \right\}.
\end{equation}
The numerical flux $\mathbf{F^{*}}$ is assumed to be the Roe flux, and an under-resolved model results in a sub-optimal  solution $\mathbf{u_h}$, i.e.,
\begin{equation}
\mathbf{u_h} \neq \mathbf{u_3} = \Pi_3 \mathbf{u},
\end{equation}
where $\Pi_3$ denotes $L_2$-projection on $\mathcal{V}_3$. Similarly, the large-scales in $\mathbf{u_h}$ will also be inconsistent i.e.   
\begin{equation}
\mathbf{\mathbf{u_{h,1}}} = \Pi_1\mathbf{\mathbf{u_h}} \neq \mathbf{u_1} = \Pi_1 \mathbf{u_3} = \Pi_1 \mathbf{u},
\end{equation}
where $\Pi_1$ is a {\color{black}$L_2$-projects} onto a coarse-space formed by tensor-product of $p=1$ basis functions in the {stream-wise} and the span-wise directions and $p=3$ basis function in the wall-normal direction. This results in coarsening in the stream-wise and the span-wise directions only. The coarse part of the solution $\mathbf{u_{h}}$ obtained after projection i.e.  $\mathbf{\mathbf{u_{h,1}}} = \Pi_1\mathbf{\mathbf{u_h}}$ is better resolved in the coarse space $\mathcal{V}_1$ in comparison to $\mathbf{u_h}$ in the fine space $\mathcal{V}_3$. This is because the numerical dissipation due to the standard numerical flux is more likely to corrupt the smaller scales in comparison to the larger, resolved scales. We can now use our model to super-resolve $\mathbf{\mathbf{u_{h,1}}}$ back to $\mathbf{u_s} \in \tilde{\mathcal{V}_3}$. as follows
\begin{equation}
\mathbf{u_s} = f_{NN}(\mathbf{\mathbf{u_{h,1}}},...),
\end{equation}
where $\mathbf{u_s}$ is the super-resolved state in the element when the coarse-scale solution in the element and its neighbours are given by $\{\mathbf{\mathbf{u_{h,1}}},...\}$. The super-resolution of each state is performed independently on 2-D planes. The size of the network used for super-resolution is reduced to 36$\times$32$\times$16$\times$16 to ensure computational efficiency. 
Finally, in a similar approach to section 6.2,  the super-resolved state is used to compute the flux terms in the DG formulation as follows:
\begin{equation}
    \int_{\Omega_k}{\partial{\mathbf{u_h}} \over \partial t}\cdot\mathbf{w_h}d\Omega-  \int_{\Omega_k}{\nabla \mathbf{w_h}: \mathbf{F(u_h)}}d\Omega + \int_{\partial \Omega_k}{ \mathbf{w_h}\cdot \mathbf{F^{*}(u_s,u_s^{-})}}d\Gamma =  0.
\label{SILES}
\end{equation}
The application of the super-resolved state $\mathbf{u_s}$ directly in the boundary flux term makes it unstable when used with the explicit R-K type time-stepping methods. To stabilize this approach (SR-LES), the super-resolution process is relaxed as follows:
\begin{align}
    \int_{\Omega_k}{\partial{\mathbf{u_h}} \over \partial t}\cdot\mathbf{w_h}d\Omega-  \int_{\Omega_k}{\nabla \mathbf{w_h}: \mathbf{F}(\mathbf{u_h})}d\Omega + \int_{\partial \Omega_k}{ \mathbf{w_h}\cdot \mathbf{F^{*}}\left((1-\lambda)\mathbf{u_h}+\lambda \mathbf{u_s},(1-\lambda)\mathbf{u_h}^{-}+\lambda \mathbf{u_s}^{-}\right)}d\Gamma =  0. \label{RSILES2} 
\end{align}
A value $\lambda=0.1-0.2$ is chosen for the following numerical simulations. Although a higher value of the relaxation factor $\lambda$ is desirable, stability generally demands $\lambda \leq 0.2$. The discretization of the viscous terms in the compressible Navier-Stokes equation is performed using the second form of Bassi and Rebay \cite{bassi2000gmres} scheme. The boundary terms arising due to the viscous fluxes are also evaluated using the under-relaxed super-resolved state similar to the inviscid fluxes.

\begin{figure}
     \centering
     \begin{subfigure}[b]{0.49\textwidth}
         \centering
         \includegraphics[width=\textwidth]{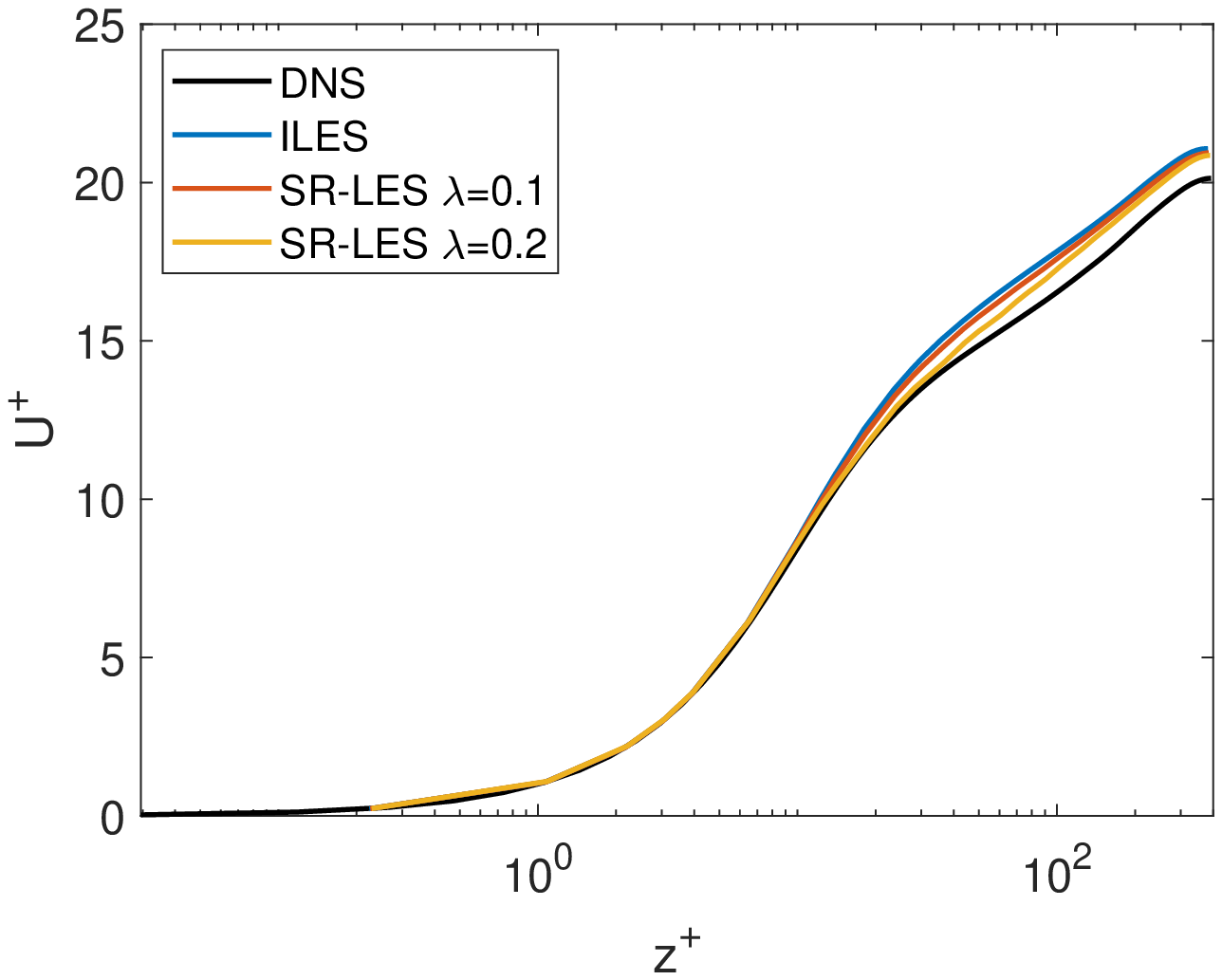}
         \caption{Stream-wise mean velocity profile $U^+$ vs. channel wall normal height $z^+$ in wall units.}
         \label{NNMEAN}
     \end{subfigure}
     \hfill
     \begin{subfigure}[b]{0.49\textwidth}
         \centering
         \includegraphics[width=\textwidth]{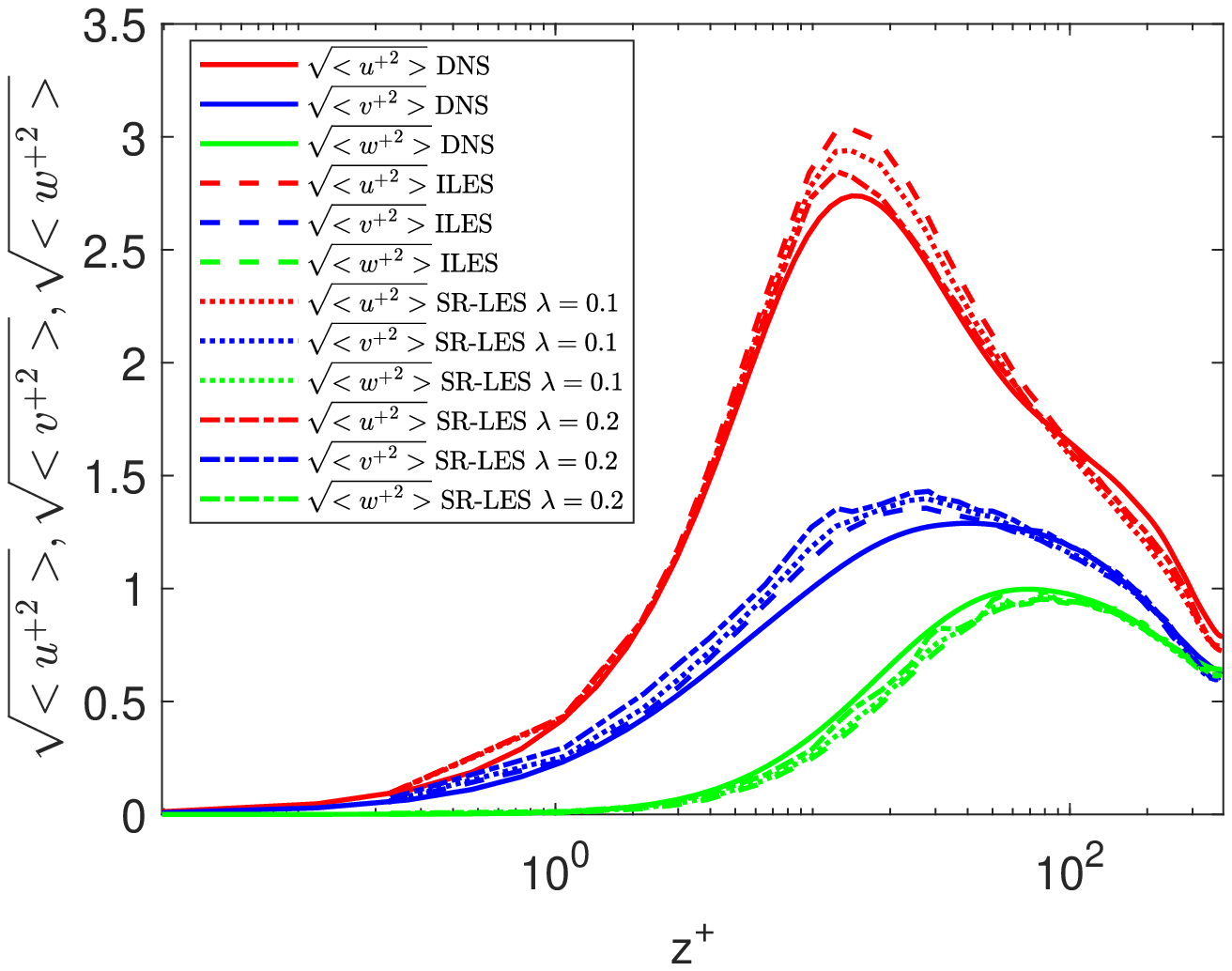}
         \caption{Root mean square of velocity components $\sqrt{<{{u^+}^2}>}$, $\sqrt{<{{v^+}^2}>}$ ,$\sqrt{<{{w^+}^2}>}$ vs. channel wall normal height $z^+$ in wall units.}
         \label{NNRMS1}
     \end{subfigure}
          \begin{subfigure}[b]{0.49\textwidth}
         \centering
         \includegraphics[width=\textwidth]{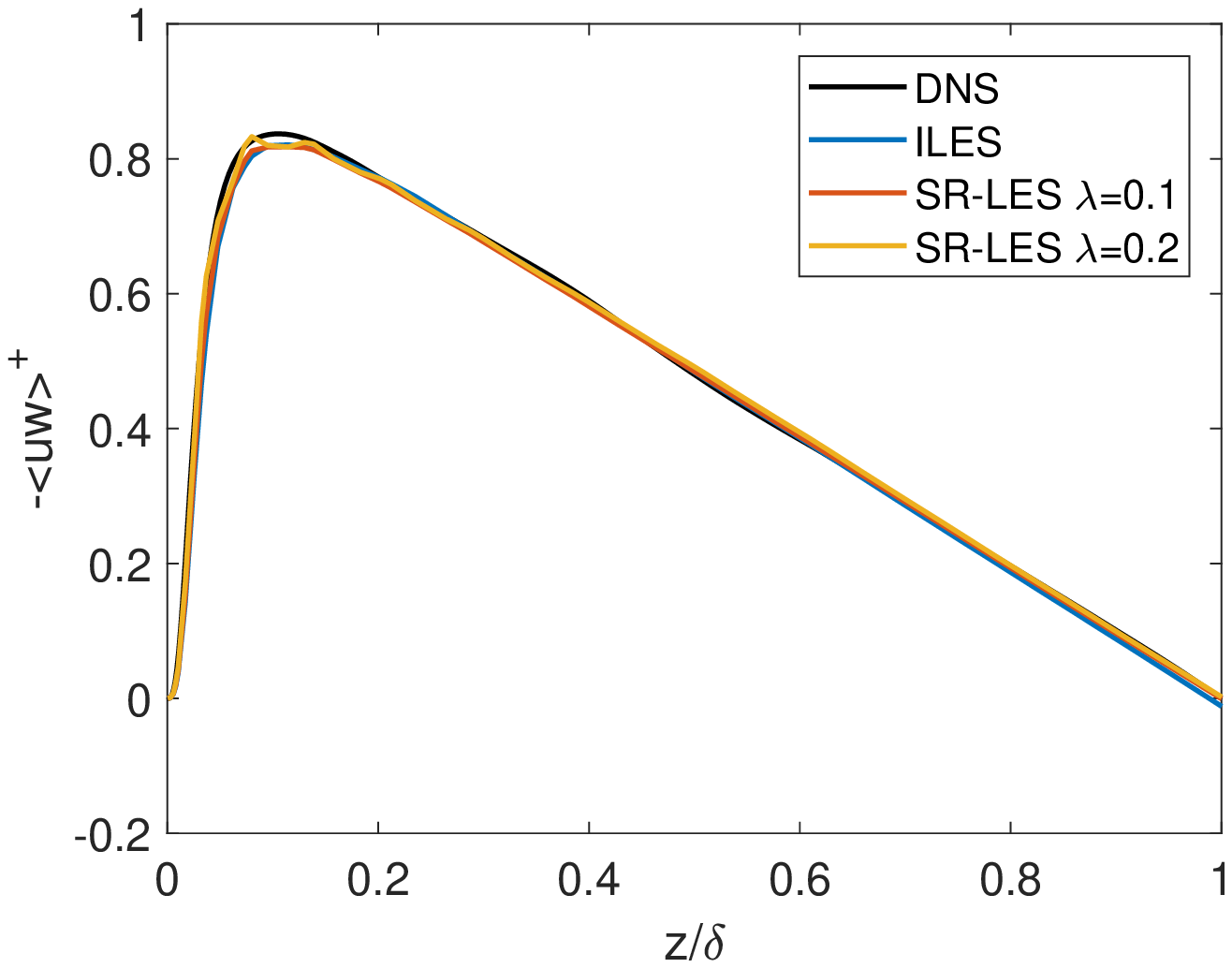}
         \caption{Resolved turbulent shear-stress -$<uw>^{+}$ vs. channel wall normal height $z/\delta$ normalised with the semi-channel height.}
         \label{NNRMS2}
     \end{subfigure}
        \caption{Velocity statistics for channel flow using ILES and SR-LES at $Re_{\tau} \approx 395$.}
        \label{fig:ILES}
\end{figure}

To compare the performance of different models, we perform LES of channel flow at $Re_{\tau} \approx 395$. The number of elements in stream-wise $(x)$, span-wise $(y)$ and wall-normal $(z)$ directions are $N_x = 24$, $N_y = 12$ and $N_z = 24$ respectively. Similarly, the size of the domain in these directions is taken to be $[L_X,L_y,L_z]:[2\pi\delta,\pi\delta,2\delta]$, respectively. The stream-wise and span-wise element sizes  in wall-units are $\Delta x^+ \approx 103.41$ and $\Delta y^+ \approx 103.41$ respectively. The element sizes in wall-normal direction vary from $\Delta z^+_{min} \approx 3.37$ near the wall to $\Delta z^+_{max} \approx 51.55$ at the center of the channel. Since, high-order basis functions are used, the effective grid size can be approximated by $\Delta_{eff} \approx {\Delta \over p}$, where $\Delta$ is the element size and $p$ is the order of the polynomial i.e. $p=3$.  Time marching was performed using the explicit RK3-TVD scheme for all the cases. Figure \ref{fig:ILES} shows the velocity statistics obtained for the channel flow problem using ILES, SR-LES and DNS. The performance improves as $\lambda$ is increased to 0.2. This is observed both in the mean velocity profile, and the stream-wise root mean square (RMS) velocity profile. The RMS peak of the stream-wise velocity obtained for both $\lambda=0.1$ and $\lambda=0.2$ is closer to DNS and lower than ILES. The span-wise and wall-normal RMS velocity statistics and the turbulent shear-stresses obtained are comparable for all three LES cases. 

A maximum relaxation factor of $\lambda=0.2$ also suggests that additional stabilization is required for the model to work at higher values of $\lambda$ where the model is expected to work better. One of the reasons for the constraint in $\lambda$ is due to the explicit time-stepping scheme was used. No such factor was needed for the linear advection case in the previous section, where an implicit space-time method was employed.  As discussed further in the perspectives section below, additional challenges have to be addressed to ensure success of super-resolution networks for predictive modeling of turbulent flows.

\section{Perspectives}

Inspired by successes in the machine vision community, there has recently been considerable interest in the use super-resolution in the physical sciences. Much of the existing literature has, however, focused on reconstruction performance and not on predictive modeling. Truly predictive models should not be restricted to a single mesh or flow configuration, and should generalize to a class of flows. Despite the success in the canonical problem in section~\ref{sec:linadv}, the results in section~\ref{sec:turbchan} suggest that there is much to be done before a truly predictive capability can be realized for a problem as challenging as turbulent flow. We view our work as a first step in moving towards a  predictive LES capability. Along these lines, we  outline the following ingredients for the discovery of sub-grid closures: 

\begin{enumerate}

\item The model should be constructed using features that lend themselves to generalization

\item The structure of the learning model should allow one to efficiently embed physics-informed parameters

\item The closure model should be intimately linked to the underlying numerical discretization. 

\item The training should be performed in manner that the super-resolution is consistent with the coarse scales during the prediction. 

\end{enumerate}

In our work, we addressed points \#1 \#2 above by by choosing non-dimensional features that are inspired by VMS closures, and by choosing a compositional neural network structure. Further work is required to design features that satisfy additional physics-informed invariances.

Regarding point \#3,  in contrast to implicitly filtered approaches in which coarse space is defined ambiguously~\cite{lund2003use}, the VMS approach formally segregates the coarse and fine spaces, thus setting a clean environment for the super-resolution. Other candidates include explicitly filtering~\cite{lund2003use} with a large test filter.

Point \#4 refers to establishing  consistency between the learning and prediction environments~\cite{duraisamy2020machine}. In essence, the training is performed on DNS data, i.e. $u^\prime = f(u_h^{DNS})$, whereas in the {\color{black}online} prediction stage, it is used as $u^\prime = f(u_h^{LES})$. As the error between the coarse scales in the LES and DNS grows, the super-resolution becomes less accurate. In other words, the  parameters of the learning model have not been inferred for {\color{black}online} performance. Model-consistent training has been successfully demonstrated in RANS closures~\cite{parish2016paradigm,singh2016using,holland2019field}, the authors are aware of only one such attempt in the context of LES~\cite{sirignano2020embedded}. However, as mentioned above, and in more detail in Ref.~\cite{sirignano2020embedded}, implicitly filtered approaches are associated with  other challenges. The VMS approach, on the other hand, allows for both numerics-consistent and model-consistent training, but the implementation of such a capability is a major undertaking is yet to be pursued by the authors in an LES context.

 As a final point, while the appeal of VMS is the segregation of scales and the prospects of deriving  closures with few phenomenological assumptions, structural models (e.g. ~\cite{pradhan2019variational}) generally perform poorly when the simulation is severely under-resolved. 
 Several attempts \cite{hughes2000large,wang2010mixed} have been made to combine traditional VMS approaches with phenomenological models like Smagorinsky in the form of mixed models. The use of data-driven techniques potentially allows us to account for these phenomenological relationships present in the data directly into the VMS model, thus, bridging the gap between phenomenological and structural modeling.

\section{Conclusions}
We proposed a strategy for multi-scale modeling in which the coarse and the fine scales are defined in terms of projection onto their respective finite element spaces, and segregated using a variational multiscale formulation. Existing variational multiscale formulations  provide guiding principles for the construction of consistent features and network architecture to define a super-resolution model of the fine scales. Particularly, we define an architecture - called the  Variational super-resolution neural network (VSRNN) - which approximates the sub-scales as a sum of products of individual functions of coarse-scales and the physics-informed parameters. This model form and network structure is inspired by  analytical expression for the sub-scales as given by the convection-diffusion equation. 
It is emphasized that traditional architectures - such as a fully connected neural network - are not ideal for this purpose because they combine heterogeneous quantities (e.g. coarse-scale basis coefficients and physics informed-parameters) as inputs. The input features and output quantities are  obtained by appropriately non-dimensionalizing the coarse-scales and the sub-scale basis coefficients. 
By applying the super-resolved state to compute the Discontinuous Galerkin (DG) fluxes, we ensure that the online coarse-scale solution is forced towards its $L_2$-optimal state.
    
We verify that when the present approach is applied to the convection-diffusion problem, it can learn the analytical solution to a high degree of accuracy. Similarly, for the 1-D linear-advection space-time problem, the model could accurately super-resolve low-order coarse solutions to high-order fine-solution. The network could also reproduce super-resolved velocity fields with the proper energy distribution across different wave-numbers in the stream-wise and the span-wise direction for the turbulent channel case. 

Next, we assessed the predictive capability of these models.  Super-resolution was used  to determine the DG fluxes for the linear-advection problem, and shown to result in higher accuracy and optimality of the method over traditional space-time methods for the same number of degrees of freedom.  When applied to LES of turbulent channel flow, this approach led to a more modest performance improvement. This improvement stems from the fact that the present model has been trained using $L_2$-optimal fine and coarse solutions, leading to sub-grid models that are consistent with the type of optimality sought. The present method was found to generalize to out-of-sample initial conditions and Reynolds numbers for both the linear advection and the turbulent channel flow cases. 
    

Perspectives were provided on data-driven closure modeling in general, and particularly how model-consistent training could improve the prospects of developing truly predictive models. In addition to reconstruction and  sub-grid modeling, the super-resolution model can be used as an error indicator for adaptive grid refinement : Regions in which the high magnitude of the sub-scale values can be used as a measure for under-resolution. Finally, the authors would like to point out that effective implementation of this approach solvers requires the development of efficient non-linear solvers and preconditioners to handle the additional non-linearity and stiffness due to the model. 
    
\section*{Acknowledgement}
This research was funded by NASA under the project "Scale-resolving turbulence simulations through adaptive high-order discretizations and data-enabled model refinements", grant number 80NSSC18M0149 (Technical monitor: Gary Coleman). We gratefully acknowledge Dr. Krzysztof Fidkowski for the valuable discussions.
{ \color{black}
\appendix
\section{'Nodally exact' high-order CG schemes for 1-D Convection-Diffusion}
To demonstrate the action of sub-scales in Section 3, we assumed that the coarse space to be composed of piece-wise linear polynomials. However, this approach can be extended to higher-order polynomials as well. In this section, we will use the VSRNN architecture to learn closures for high order CG discretizations where the coarse-scale is 'nodally exact'. The governing PDE is again taken to be the linear convection-diffusion equation as follows:
\begin{equation}
    \mathcal{L} \triangleq a \frac{d}{dx} - {\kappa}\frac{d^2}{dx^2} \quad in \quad \Omega=[0,L]
\end{equation}
with Dirichlet boundary conditions: $u(0) = u_0$ and $u(L) = u_L$. To derive VSRNN closures for 'nodally exact' coarse-scales, we start with the variational form:
\begin{equation}
(a{d u \over d x}-\nu {d^2 u \over d x^2},w)=0
\end{equation}
The weak form after integration by parts is obtained as follows:
\begin{equation}
\left(a \frac{d u}{d x}, w\right)+\kappa\left(\frac{d u}{d x}, \frac{d w}{d x}\right)=0
\end{equation}
The next step is to apply the VMS decomposition such that the coarse-scale is exactly the interpolate of the true solution at nodal points i.e.
\begin{equation}
( a{d \tilde{u} \over d x},\tilde{w}) + \kappa ({d \tilde{u} \over d x},{d \tilde{w} \over d x}) + ( a{d {u'} \over d x},\tilde{w}) + \kappa ({d {u'} \over d x},{d \tilde{w} \over d x}) = 0
\end{equation}
Since the coarse-scale is the true interpolant of the solution, the sub-scales should vanish at the nodal points. Hence, integration by parts can be performed as follows:
\begin{equation}
( a{d \tilde{u} \over d x},\tilde{w}) + \kappa ({d \tilde{u} \over d x},{d \tilde{w} \over d x}) + \sum_{e}\int_{\Omega_e} u' (-a{d \tilde{w} \over d x}-\kappa{d^2{\tilde w}\over{d x^2}}) d\Omega = 0
     \label{VMS_CD_P}
\end{equation}
It can be recognized that the sub-scale lies in an infinite dimensional space. However, for each element only its inner-product with $-a{d \tilde{w} \over d x}-\kappa{d^2{\tilde w}\over{d x^2}}$ needs to be computed inside each element. Hence, if $p$ is the order of the polynomial used to describe the coarse-scales, $L_2$-projecting $u'$ in a discontinuous polynomial space inside the element consisting of polynomials up to order $p-1$ is sufficient. To learn these projected sub-scales, the VSRNN is used as follows:
\begin{align}
    \left[C'_{1,p-1},C'_{2,p-1},..,C'_{p,p-1}\right] = \mathbf{f}\biggr(log(\alpha),\left[\tilde{C}_{1,p},\tilde{C}_{2,p},..,\tilde{C}_{p+1,p}\right]\biggr), \nonumber
\end{align}
where, $\left[C'_{1,p-1},C'_{2,p-1},..,C'_{p,p-1}\right]$ represents the sub-scale basis coefficients normalised by coarse-scale R.M.S $u_{rms}$, $[\tilde{C}_{1,p},\tilde{C}_{2,p},..,\tilde{C}_{p+1,p}]$ represents the mean $u_{mean}$ subtracted and $u_{rms}$ normalised coarse-scale basis coefficients, and $\alpha$ is the Peclet number. 

It is noted that although $u'$ is infinite dimensional, $\left[C'_{1,p-1},C'_{2,p-1},..,C'_{p,p-1}\right]$ is not. It is sufficient to learn these projected sub-scales to precisely compute the required inner-products. For example, when the coarse solution is linear ($p=1$), the sub-scale can be represented by $p=0$ constant functions which corresponds to commonly used $\tau = {h \over 2 a}({coth(\alpha)-{1\over \alpha}})$. The online evaluation of the $p=1$ closure presented in Figure \ref{fig:tau0} is shown in figure \ref{fig:CDNN1}. As expected, the VSRNN is able to precisely recreate the results obtained using the analytical expression for $\tau$. 
\begin{figure}
     \centering
     \begin{subfigure}[b]{0.40\textwidth}
         \centering
         \includegraphics[width=\textwidth]{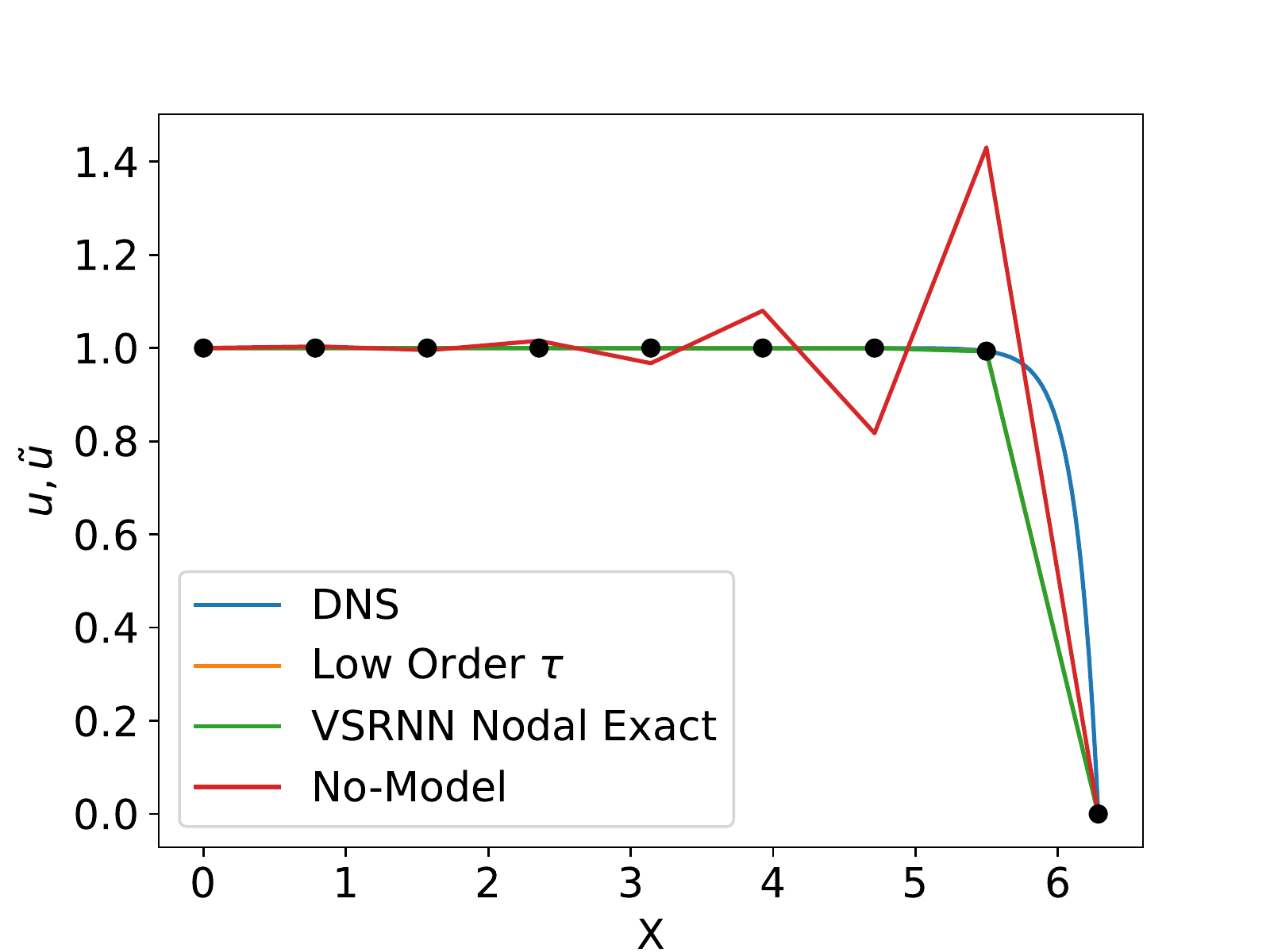}
         \caption{$Pe_g = 20$.}
         \label{CDNN1_20}
     \end{subfigure}
     \begin{subfigure}[b]{0.40\textwidth}
         \centering
         \includegraphics[width=\textwidth]{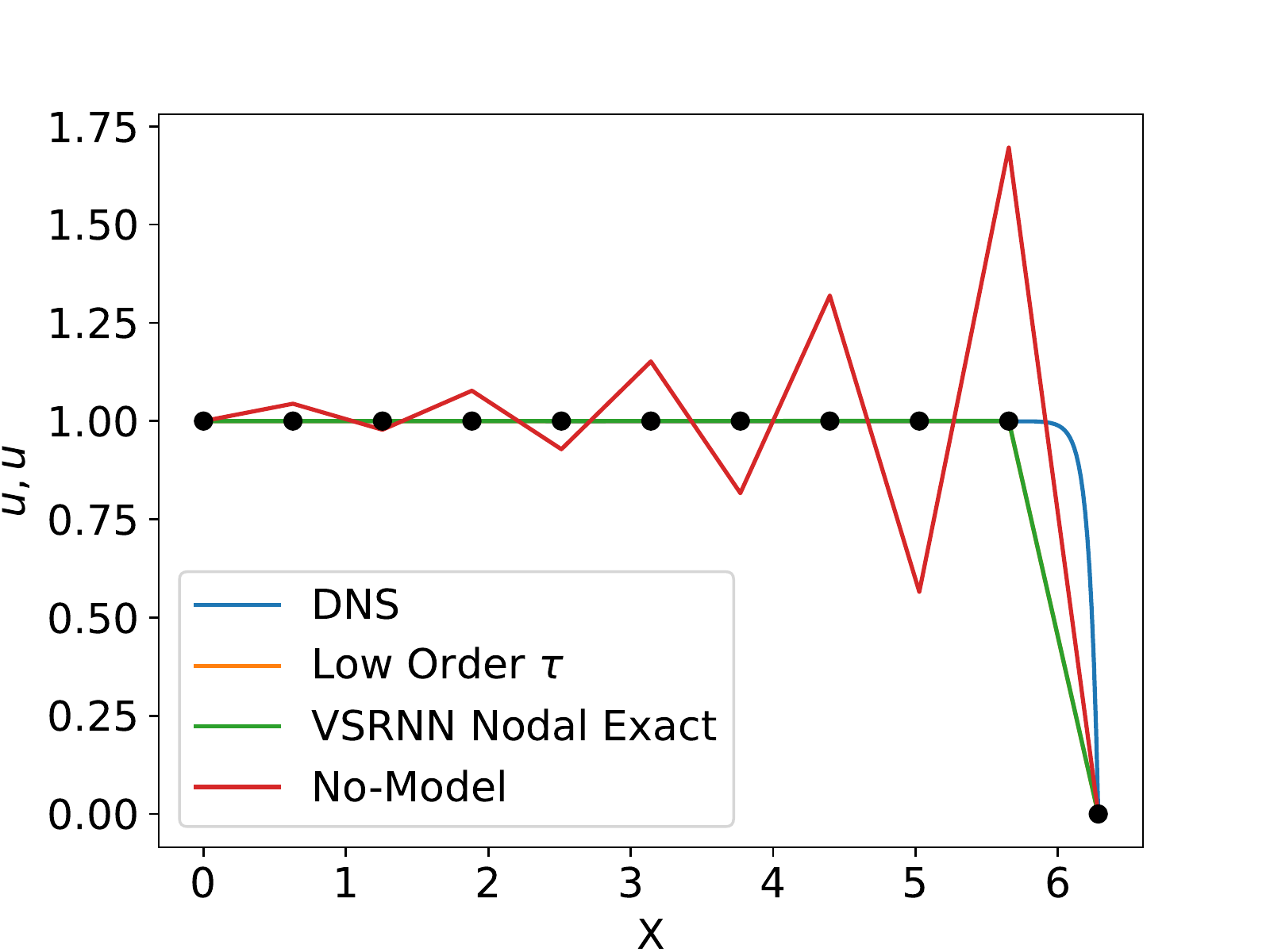}
         \caption{$Pe_g = 40$.}
         \label{CDNN1_40}
     \end{subfigure}
    \caption{Comparison of VSRNN closure to existing closure for $p=1$ CG finite elements at different Peclet numbers.}
    \label{fig:CDNN1}
\end{figure}

The data generation procedure used here for deriving closures for higher order polynomials is same as that used in Section 4 for the $p=1$ case. The model is trained and applied to equation \eqref{VMS_CD_P}. Two different cases with global Peclet numbers $Pe_{g} = {{a L}\over{\kappa}}=$20 and 40 are considered here. For each case, two CG elements with different polynomial orders $p=3,4,7$ and $8$ are used to discretize the domain. Figure \ref{fig:CDNN} and \ref{fig:CDNN2} shows the comparison of the present VSRNN closure to existing closures and no-model for global Peclet numbers of 20 and 40 respectively. The VSRNN model in both the cases accurately learns the sub-scales and ensures that the coarse-scale is the interpolate of the true solution. As expected, $\tau = {h \over 2 a}({coth(\alpha)-{1\over \alpha}})$ based on low-order discretization is not accurate at high-orders and the no-model discretization (Galerkin) is oscillatory when resolution is not sufficient. The no-model performance increases at high-order because effective resolution increases with $p$.

\begin{figure}
     \centering
     \begin{subfigure}[b]{0.40\textwidth}
         \centering
         \includegraphics[width=\textwidth]{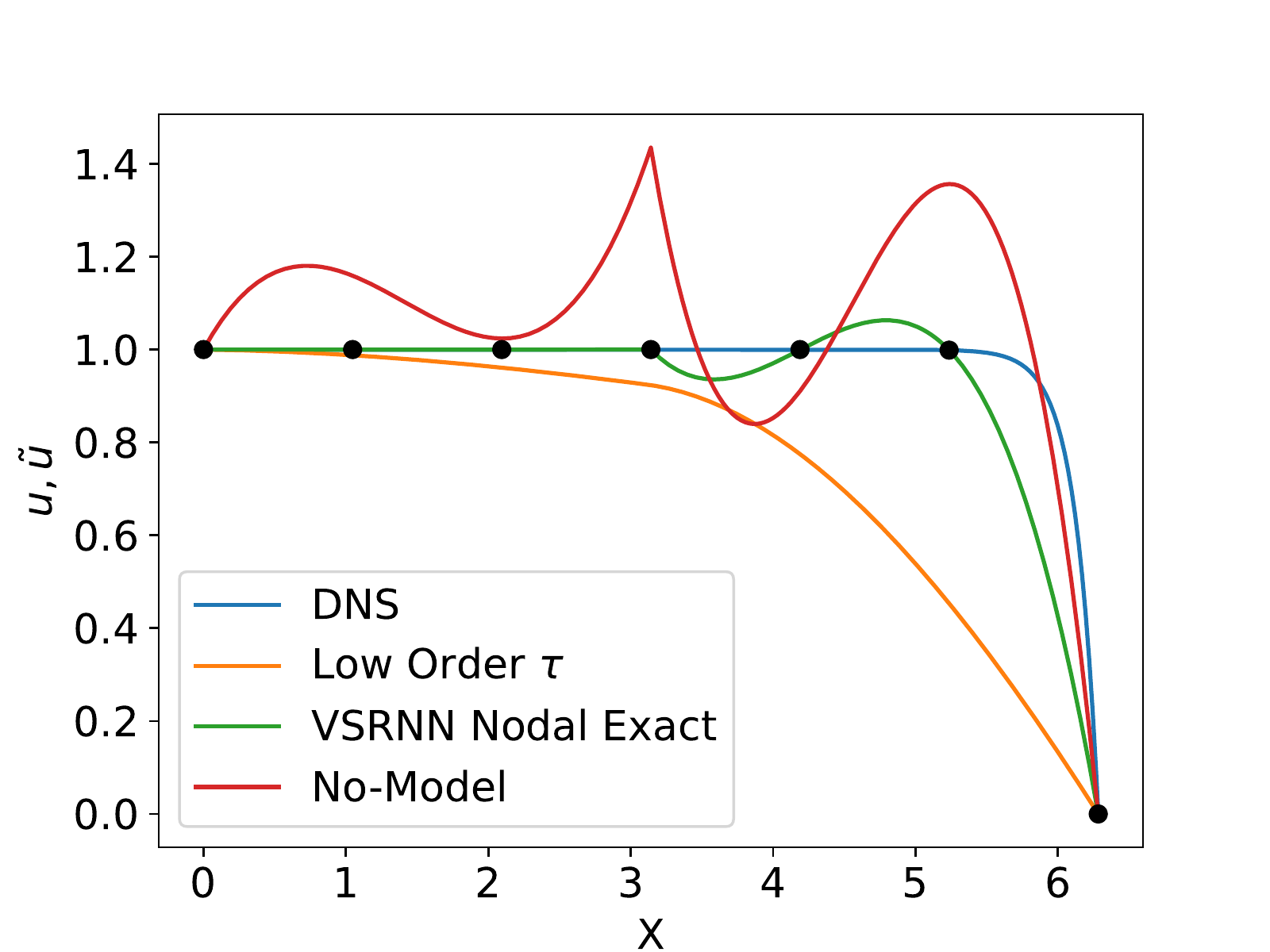}
         \caption{$p=3$.}
         \label{CDNN3}
     \end{subfigure}
     \begin{subfigure}[b]{0.40\textwidth}
         \centering
         \includegraphics[width=\textwidth]{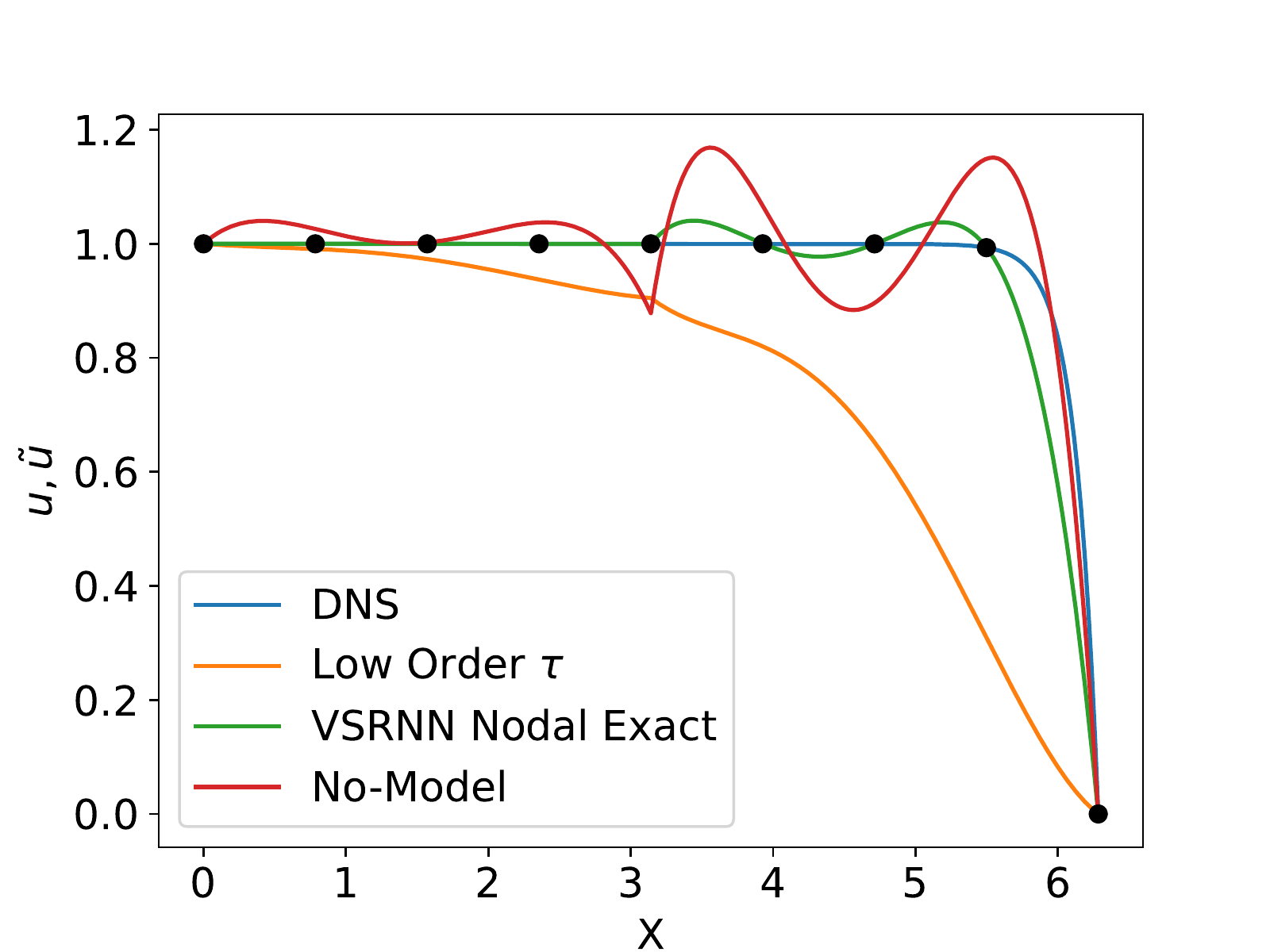}
         \caption{$p=4$.}
         \label{CDNN4}
     \end{subfigure}
     \begin{subfigure}[b]{0.40\textwidth}
         \centering
         \includegraphics[width=\textwidth]{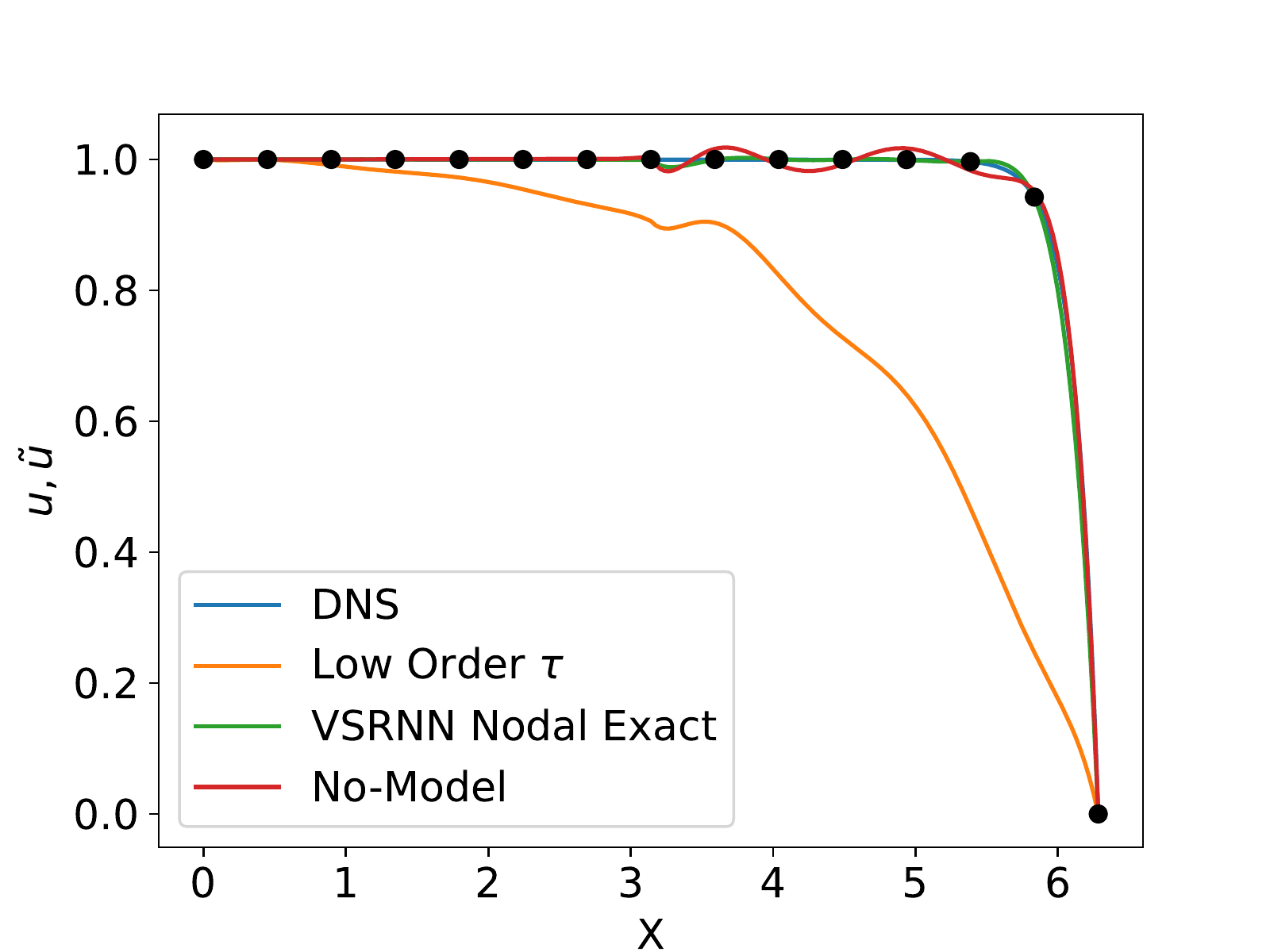}
         \caption{$p=7$.}
         \label{CDNN7}
     \end{subfigure}
     \begin{subfigure}[b]{0.40\textwidth}
         \centering
         \includegraphics[width=\textwidth]{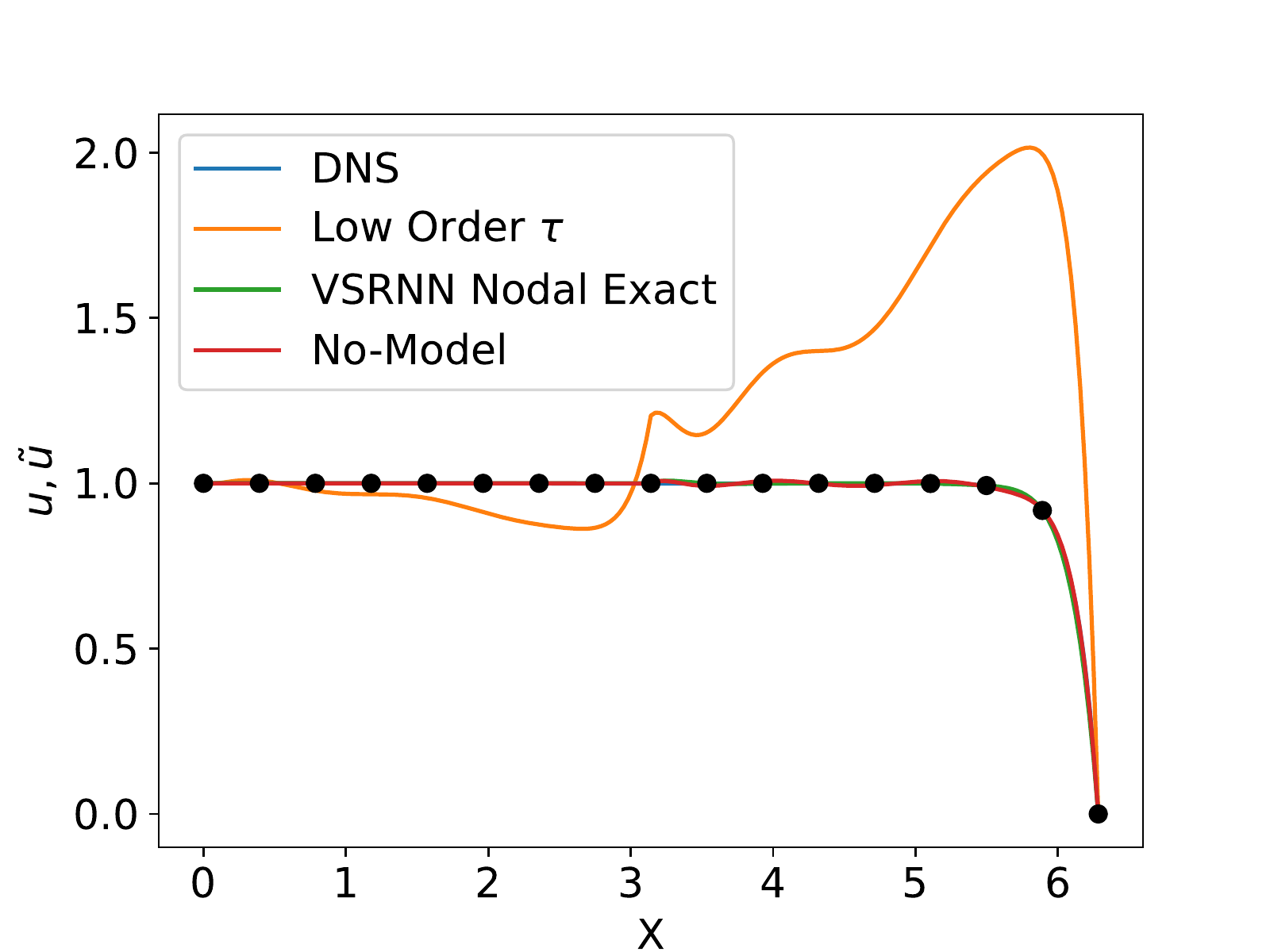}
         \caption{$p=8$.}
         \label{CDNN8}
     \end{subfigure}
        \caption{Discretizations of the 1-D Convection-Diffusion equation at $Pe_{g} = 20$ using two CG elements $N_{ele}=2$ and different polynomial orders $p=3,4,7,8$.}
        \label{fig:CDNN}
\end{figure}

\begin{figure}
     \centering
     \begin{subfigure}[b]{0.40\textwidth}
         \centering
         \includegraphics[width=\textwidth]{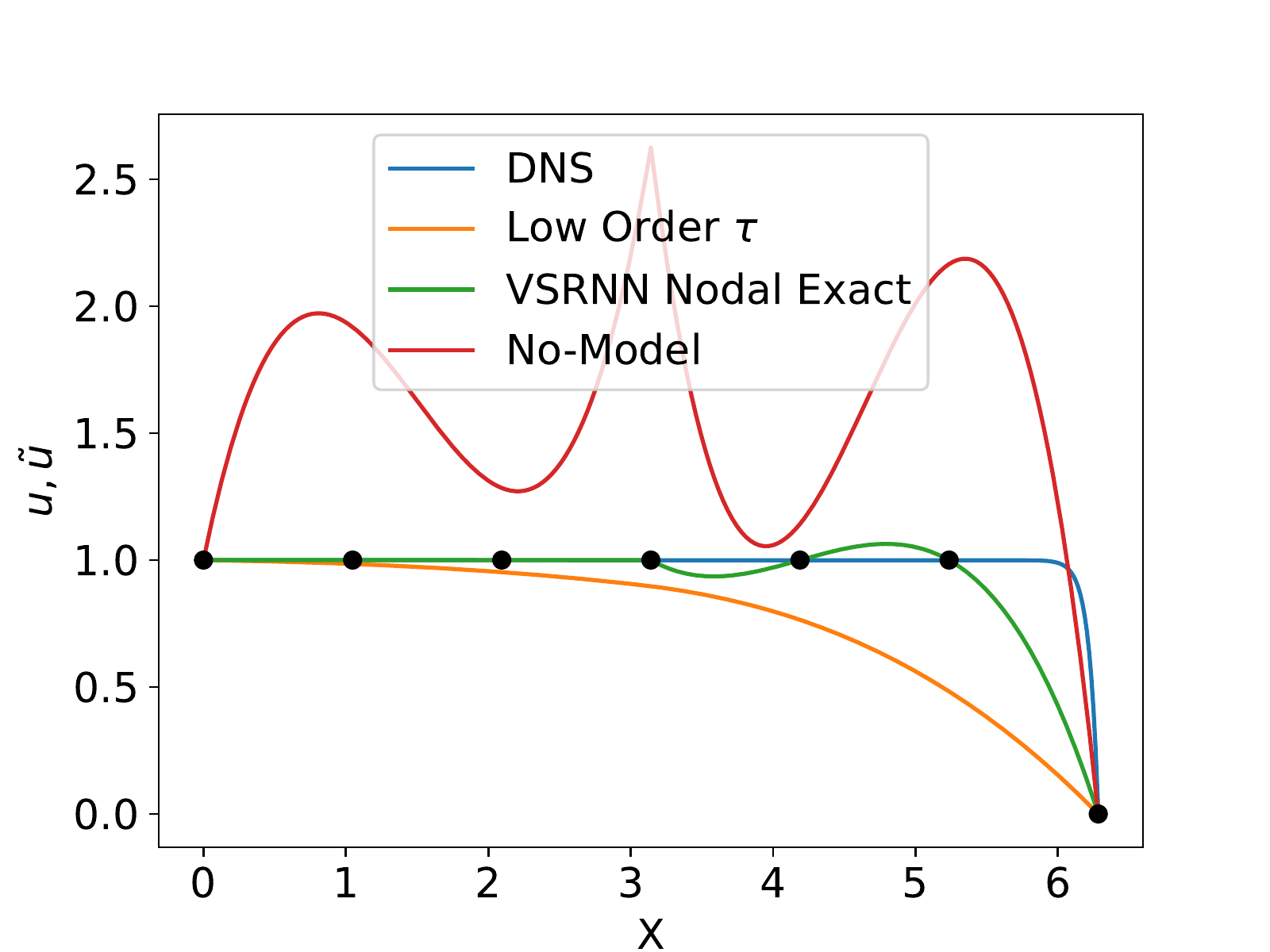}
         \caption{$p=3$.}
         \label{CD2NN3}
     \end{subfigure}
     \begin{subfigure}[b]{0.40\textwidth}
         \centering
         \includegraphics[width=\textwidth]{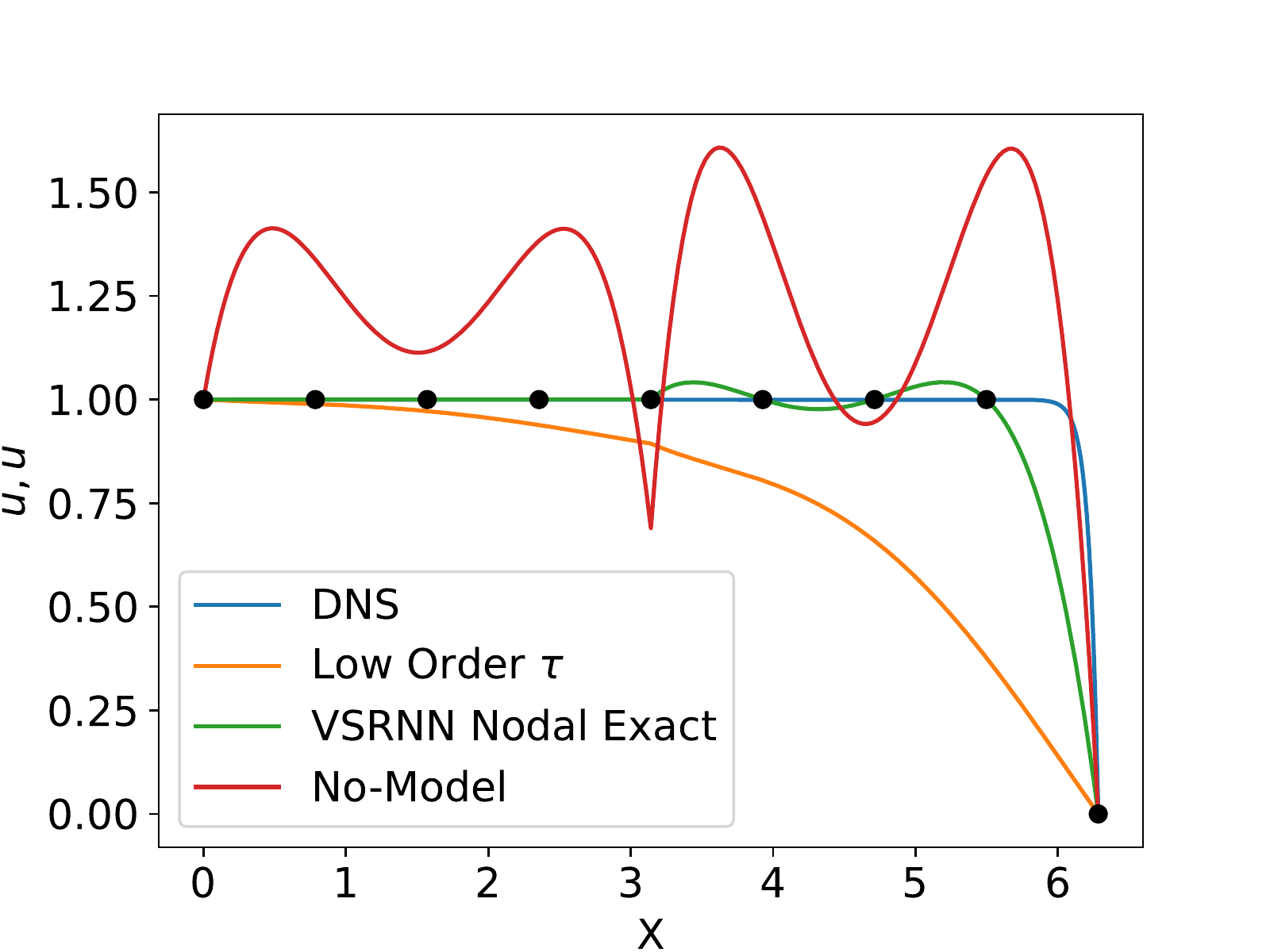}
         \caption{$p=4$.}
         \label{CD2NN4}
     \end{subfigure}
     \begin{subfigure}[b]{0.40\textwidth}
         \centering
         \includegraphics[width=\textwidth]{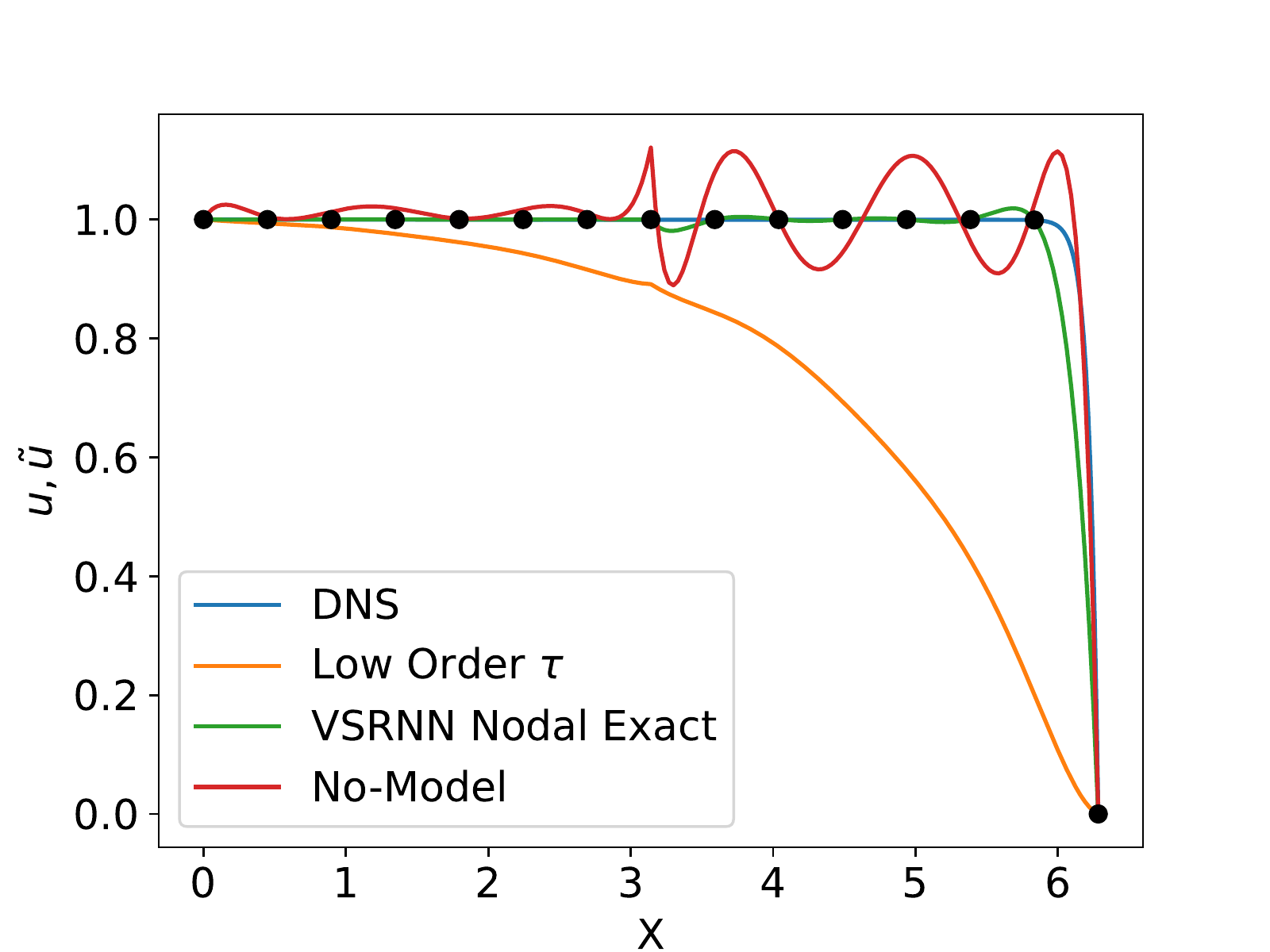}
         \caption{$p=7$.}
         \label{CD2NN7}
     \end{subfigure}
     \begin{subfigure}[b]{0.40\textwidth}
         \centering
         \includegraphics[width=\textwidth]{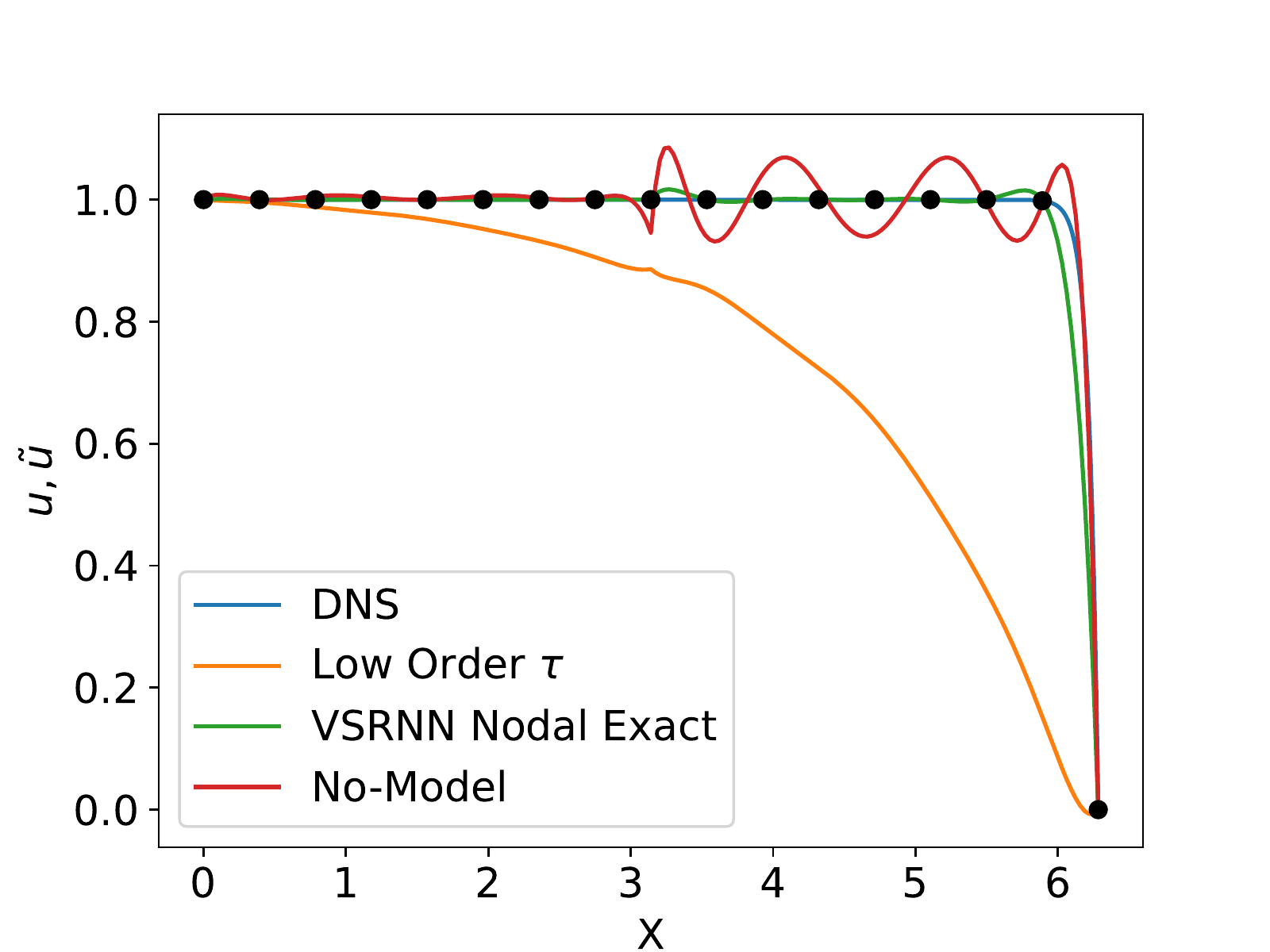}
         \caption{$p=8$.}
         \label{CD2NN8}
     \end{subfigure}
        \caption{Discretizations of the 1-D Convection-Diffusion equation at $Pe_{g} = 40$ using two CG elements $N_{ele}=2$ and different polynomial orders $p=3,4,7,8$.}
        \label{fig:CDNN2}
\end{figure}
}

\bibliography{mybibfile}

\end{document}